\documentclass{jfm}
\usepackage{graphicx}
\usepackage{epstopdf, epsfig}
\usepackage{amsmath}
\usepackage{amssymb}
\usepackage[colorlinks, citecolor=blue]{hyperref}
\usepackage[usenames,dvipsnames]{xcolor}
\usepackage{bigints}
\usepackage{longtable}
\usepackage{colortbl}
\usepackage{booktabs}
\usepackage[font=small]{caption}
\newcommand\hyd{\mathit{\mathrm{H_2}}}
\newcommand\acid{\mathit{\mathrm{H_2}\mathrm{SO_4}}}
\newcommand\hion{\mathit{\mathrm{H^+} } }
\newcommand\sulphate{ \mathit{ \mathrm{SO^{2-}_4} } }
\newcommand\Shge{\mathit{Sh_{\mathrm{\, H_2,e}} }}
\newcommand\Shget{\mathit{\tilde{Sh}_{\mathrm{\, H_2,e}} }}
\newcommand\Shgei{\mathit{Sh^{-\textrm{1}}_{\mathrm{\, H_2,e}}}}
\newcommand\Shgb{\mathit{Sh_{\mathrm{\, H_2,b}}}}
\newcommand\Shgbt{\mathit{\tilde{Sh}_{\mathrm{\, H_2,b}}}}
\newcommand\Shgbs{\mathit{Sh^\ast_{\mathrm{\, H_2,b}}}}
\newcommand\Shs{\mathit{Sh_{\mathrm{\, s,e}}}}
\newcommand\Shst{\mathit{\tilde{Sh}_{\mathrm{\, s,e}}}}
\newcommand\Shj{\mathit{Sh_{\, j,\mathrm{e}}}}
\newcommand\Shjt{\mathit{\tilde{Sh}_{\, j,\mathrm{e}}}}
\newcommand\Pes{\mathit{Pe^\ast}}
\newcommand\Jas{\mathit{Ja^\ast}}
\newcommand\Gr{\mathit{Gr}}
\newcommand\Scg{\mathit{Sc_{\mathrm{\, H_2}}}}
\newcommand\Scs{\mathit{Sc_{\mathrm{\, s}}}}
\newcommand\Scj{\mathit{Sc_{\, j}}}
\newcommand\Ja{\mathit{Ja}}
\newcommand\Csath{\mathit{C_{\mathrm{\, H_2,sat}}}}
\newcommand\Cg{\mathit{C_\mathrm{H_2,e}}}

\newcommand\Cst{\mathit{\tilde{C}_\mathrm{s,e}}}
\newcommand\Cgt{\mathit{ \tilde{C}_\mathrm{H_2,e}}}
\providecommand\dd{\mathrm{d}}

\definecolor{Gray}{gray}{0.85}

\shorttitle{Mass transport at gas-evolving electrodes}
\shortauthor{ Sepahi, Verzicco, Lohse, Krug}

\title{Mass transport at gas-evolving electrodes}

\author{Farzan Sepahi\aff{1}
	\corresp{\email{f.sepahi@utwente.nl}},
	Roberto Verzicco\aff{1,3,4}
	Detlef Lohse\aff{1,2}
	\and Dominik Krug\aff{1}
	\corresp{\email{d.j.krug@utwente.nl}}}

\affiliation{\aff{1}Physics of Fluids group, Max Planck Center for Complex Fluid Dynamics, MESA+ Institute and J. M. Burgers Centre for Fluid Dynamics, University of Twente, P.O. Box 217, 7500AE Enschede, Netherlands
	\aff{2}Max Planck Institute for Dynamics and Self-Organization, Am Fassberg 17, 37077 Göttingen, Germany
	\aff{3}Dipartimento di Ingegneria Industriale, University of Rome ‘Tor Vergata’, Via del Politecnico 1, Roma 00133, Italy
	\aff{4}Gran Sasso Science Institute, Viale F. Crispi, 7 67100 L'Aquila, Italy}

\begin{document}
	\maketitle

	\begin{abstract}
		Direct numerical simulations are utilised to investigate mass transfer processes at gas-evolving electrodes that experience successive formation and detachment of bubbles. The gas-liquid interface is modeled employing an Immersed Boundary Method. We simulate the growth phase of the bubbles followed by their departure from the electrode surface in order to study the mixing induced by these processes.  We find that the growth of the bubbles switches from a diffusion-limited mode at low to moderate fractional bubble-coverages of the electrode to reaction-limited growth dynamics at high coverages. Furthermore, our results indicate that the net transport within the system is  governed by the effective buoyancy driving induced by the rising bubbles and that mechanisms commonly subsumed under the term `microconvection' do not significantly affect the mass transport. Consequently, the resulting gas transport for different bubble sizes, current densities, and electrode coverages can be collapsed onto one single curve and only depends on an effective Grashof number. The same holds for the mixing of the electrolyte when additionally taking the effect of surface blockage by attached bubbles into account. For the gas transport to the bubble, we find that the relevant Sherwood numbers also collapse onto a single curve when accounting for the driving force of bubble growth, incorporated in an effective Jakob number. Finally, linking the hydrogen transfer rates at the electrode and the bubble interface, an approximate correlation for the gas-evolution efficiency has been established. Taken together, these findings enable us to deduce parametrizations for all response parameters of the systems. 
	\end{abstract}


	\section{Introduction}\label{sec:Intro}
	\noindent Production of green hydrogen through water electrolysis is projected to be an important technology to cope with the volatile output from renewable power sources in the future energy mix and as a sustainable feedstock in various industrial processes \citep{turner2004,Holladay2009,Nikolaidis2017,Dawood2020}. For the required upscaling of the production, the formation of gas bubbles on the electrode surface plays a critical role. Attached bubbles lower the efficiency of the electrolyser systems by blocking the active electrode area \citep{Qian1998,Vogt2005,Swiegers2021}. In addition, they increase the cell resistance by lowering the effective conductivity of the electrolyte \citep{Zhao2019,Dukovic1987,Darband2019} which leads to cell overpotential. However, the formation of bubbles is also beneficial as it enhances the mixing of the electrolyte and this aspect will be the main focus of this work.
 
    The evolution of bubbles comprises nucleation, growth, and detachment from the electrode surface. Bubble growth occurs due to the diffusive transport of dissolved hydrogen to the gas-liquid interface and its subsequent desorption to the gas phase \citep{Rousar1975,Angulo2020}. The eventual detachment may be buoyancy driven \citep{Fritz1935,Slooten1984} but can also be a consequence of coalescence events \citep{Iwata2022}. Bubble evolution can impact mass transfer at the electrode in several ways. This includes local `micro-convection' and diffusion processes induced by bubble growth and break-off from the electrode surface \citep{Stephan1979,Vogt2015}, and also `macro-convection' within the bulk electrolyte caused by frequent detachment and rise of bubbles within the electrolyte solution \citep{Janssen1979,Boissonneau2000,Vogt2011b,Taqieddin2017}. The latter process is also referred to as two-phase buoyancy-driven convection as it is resulting from the density variations in gas-in-liquid dispersion, and enhances the mass transport by mixing the electrolyte solution in electrode proximity via the established macro-flow pattern. Similar to forced convection effects induced by pressure gradient or magnetic field \citep{Iida2007,Koza2011,Matsushima2013,Baczyzmalski2016,Baczyzmalski2017}, such flow structure pumps the fresh bulk electrolyte to the electrode surface replacing the reactant-depleted and gas-enriched solution in the electrode boundary layer \citep{Zuber1963}. The significance of two-phase buoyancy-driven convection is further emphasized by the fact that the efficiency of electrochemical systems reduces remarkably under microgravity condition. This adverse effect was attributed to the prolonged adherence of the bubbles to the electrode, inhibiting proper mixing, as well as their growth to inordinate sizes, which further impeded the mass transfer to the electrode \citep{Iwasaki1997,Matsushima2003,Matsushima2009,Mandin2014,Sakuma2014,Bashkatov2021}.  
	
	These different mass transfer mechanisms were studied separately in the literature. \citet{Ibl1971} established the first mass transfer relation for the diffusive micro-processes associated with bubble evolution. This model neglected convection and focused on reactant diffusion to a microarea on the electrode surface affected during the waiting period after bubble detachment and nucleation of the subsequent one. This relation was later modified by \citet{Rousar1975} and \citet{Vogt2015} to additionally account for diffusive transport during bubble growth, when the size of the microarea shrinks over time and becomes inactive under the bubble foot. 
 
    The impact of micro-convection resulting from bubble growth on mass transfer at the microarea, was first quantified by \cite{Stephan1979}. Later, \citet{Vogt2015} also took the effect of the wake into consideration, which is induced by the bubble break-off, on mass transfer at the microarea. Based on their considerations, these authors concluded that micro-convection of bubble growth and detachment is the primary controlling factor for mass transfer when the gas-evolution rate is sufficiently high, particularly at moderate and large current densities. This model is almost exclusively based on theoretical considerations, but has extensively been used for practical applications by other authors \citep{Burdyny2017,Yang2019}.
	
	In contrast to the findings of \citet{Stephan1979} and \citet{Vogt2015}, who identified the micro-convective processes of gas-evolution as the dominant mechanism, \citet{Janssen1970,Janssen1973,Janssen1978,Janssen1979} provided evidence that mass transfer at the electrode was governed by two-phase free convection driven by rising bubbles. This was corroborated by measurements conducted on hydrogen evolving electrodes, with no coalescence of bubbles, where the boundary layer thickness, as a function of volumetric gas evolution rate, exhibited a power law relationship with an exponent of $1/3$. This observation highlighted the analogy between such flows, induced by density variations in gas-in-liquid dispersion, and single-phase natural convection in heat and mass transfer problems \citep{Churchill1975,Wragg1968}. 
 
    In summary, the findings by different authors on the relevance of the various transport processes close to the gas-producing electrodes are contradictory, and as of our current knowledge, there is no consensus on the rate-controlling mechanism, let alone a well-controlled quantification, of mass transfer at gas-evolving electrodes.
	
	Numerous attempts have been made in the literature to combine experiments and numerical simulations to study the bubble-induced convection at gas-evolving electrodes \citep{Hreiz2015a}. Hydrodynamics of two-phase flow and their influence on mass transfer and reaction rate at the electrode have been modeled employing Euler-Euler \citep{Schillings2015,Abdelouahed2014a,Abdelouahed2014b,Zarghami2020,Obata2020,Obata2021}, or Euler-Lagrange \citep{Mandin2005,Hreiz2015b,Hreiz2015a,Battistella2018} approaches, in neither of which the gas-liquid interface of the bubble were resolved. However, only interface-resolved simulations are capable of capturing the micro-convection as a result of bubble growth and break-off. Several authors performed numerical simulations to study the dynamics of bubble growth coupled with electrokinetics of gas-evolution reaction at the electrode using immersed boundary method (IBM) \citep{Khalighi2023} or body-fitted grids \citep{Higuera2021,Higuera2022}. Other relevant dynamics of bubbles near the electrodes such as coalescence, detachment and rising have separately been investigated with interface-resolved simulations \citep{Zhang2020,Torii2021}. However, none of these studies simultaneously treat the effect of bubble growth induced micro-convection and two-phase buoyancy-driven convection.  
	
	Despite numerous studies targeting the interplay between two-phase hydrodynamics and electrochemical phenomena at gas-evolving electrodes, the question of whether the primary mass transfer mechanism is attributed to the micro-convective processes of bubble growth \citep{Stephan1979,Vogt2015} or two-phase free convection of gas-in-liquid dispersion \citep{Janssen1979}, remains unsettled. Therefore, we aim to perform interface-resolved direct numerical simulations to account for various mechanisms in play by electrolytically-generated gas bubbles. In particular, we look into the successive processes of bubble growth and rise in the electrolyte solution \citep{VanDerLinde2017,Raman2022} until an equilibrium state is reached, i.e. the global statistics of the system no longer varies in time. Our findings provide a broader perspective on the different mass transfer processes at the electrode and bubble interface by leveraging disentangled parameters in the numerical simulations.  
	
	The remainder of this paper is structured as follows; the problem set-up and governing equations are discussed in $\S$~\ref{sec:problem}. The results for bubble dynamics and mass transfer rates at the electrode are presented in $\S$~\ref{sec:electrode}. Mass transfer to the bubble and gas-evolution efficiency are quantified in $\S$~\ref{sec:Sh_bub} and $\S$~\ref{sec:fG}. Finally, we further discuss and summarize our findings in $\S$~\ref{sec:conclusion}.

	\begin{figure}
		\centering
		\includegraphics[width=1\columnwidth]{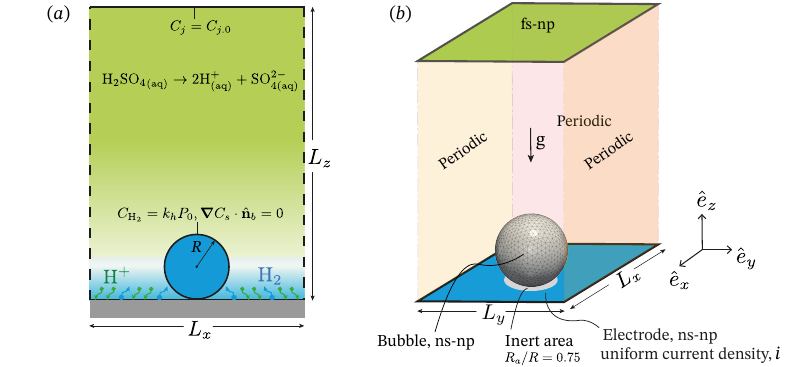}
		\caption{($a$) Schematic representation of the two-phase electrochemical system with relevant chemical reactions and boundary conditions at the cathode.  ($b$) Sketch of the 3-dimensional numerical setup with the applied boundary conditions for the velocity field (periodic, no-slip (ns), no-penetration (np) and free-slip (fs)). The bubble is modeled with IBM using a triangulated Lagrangian grid on the bubble interface (a sample is illustrated in panel ($b$)). Current density is uniformly distributed on the electrode surface except for an inactive $(i=0)$ circular part with an outer radius of $R_a=0.75R$ under the bubble.}
		\label{fig:Numerical_sketch}
	\end{figure}

	\section{Configuration and numerical methods}\label{sec:problem}
    \subsection{Problem set-up}
	\noindent The electrochemical model considered here concerns a water-splitting system with dilute sulfuric acid ($\acid$, 500~$\mathrm{mol/m^3}$) as electrolyte. A schematic is provided in figure \ref{fig:Numerical_sketch}($a$) demonstrating the chemical reactions at the cathodic part of the cell.  Full dissociation of sulfuric acid to sulphate ($\sulphate$) and hydrogen ($\hion$) ions is assumed according to 
	
	\begin{align}\label{eq:AciDdissociation}
		\mathrm{H_2SO_{4(aq)}} \rightarrow 2\mathrm{H^{+}_{(aq)}} + \mathrm{SO_{4(aq)}^{2-}},
	\end{align} 
	
	\noindent and in order to avoid further complications, self-ionization of water is disregarded due to its low equilibrium constant at room temperature.  The cathodic reactions solely comprise the Hydrogen Evolution Reaction (HER) as 
	
	\begin{align}\label{eq:HER}
		2\hion + 2\mathrm{e^-} \rightarrow \hyd,
	\end{align} 
	
	\noindent whereby the hydrogen enrichment and electrolyte depletion co-occur within a mass-transfer boundary layer in the vicinity of the electrode as schematically illustrated in figure \ref{fig:Numerical_sketch}($a$).  
	
	The numerical set-up is a cuboid box as depicted in figure \ref{fig:Numerical_sketch}($b$). The electrode is oriented horizontally ($x$ and $y$ directions) such that the gravitational acceleration $\mathbf{g}$ acts normal to it in the negative $z$ direction. A fully spherical hydrogen bubble is initialized with a certain radius ($R_0=50~\mu$m) and zero-degree contact angle on the electrode. The bubble subsequently grows to a prescribed diameter, namely the break-off diameter $d_b$, before it departs from the electrode surface and rises within the electrolyte solution due to its buoyancy. This process then repeats with the next bubble initialized at the same spot as soon as the previous bubble exits from the top boundary. By applying periodicity in the lateral directions of the computational domain, the set-up replicates a system of monodisperse bubbles with uniform spacing of $S=L_x=L_y$, which synchronously grow and rise in the medium. The initialization, growth, and rise of the bubbles in succession are modeled until an equilibrium state is attained, i.e. the averaged mass transfer statistics, which will be introduced in $\S$ \ref{sec:response_param}, remain constant in time. 
	
	The control parameters for the electrolytically-generated two-phase free convective flow are the cathodic current density $i$, the bubble break-off diameter $d_b$, and the bubble spacing $S$. Simulations are performed with two different sets of configuration as listed in table \ref{Table1_ch2}; in the first set the bubble spacing is kept constant while the bubble break-off diameter is varied. In the second set, spacing between the bubbles is varied at a constant break-off diameter of the bubbles to investigate the effect of bubble population density on the mass transport at the electrode. An auxiliary parameter for either set is the fractional bubble coverage of the electrode, $\Theta$, which refers to the fraction of the electrode area shadowed by the orthogonal projection of the bubble surface. It is formulated as $\Theta=\pi d_b^2/4A_e$, where $A_e=L_xL_y$ is the electrode area available for a single bubble. At each configuration, 13 current densities within the range $10^1 \le \vert i \vert \le 10^4~\mathrm{A/m^2}$, as listed in table \ref{Table1_ch2}, are simulated. 
	
\begin{table}
\tabcolsep 18 pt
\begin{tabular}{ccccc}
\multicolumn{4}{c}{Configuration}   &    $\vert i \vert~\mathrm{[A/m^2]}$\\
\cmidrule(lr){1-4}\cmidrule(lr){5-5}
\rowcolor{Gray}
     &  \multicolumn{3}{c}{Constant bubble spacing, $L_x=L_y=2$ mm }   &   \cellcolor{White} $10^1$\\
No.    &   $d_b$ [mm]  &  $\Theta$    &   $N_x \times N_y \times N_z$ &   $1.7 \times 10^1$\\
\cmidrule(lr){1-4}
1   &   0.3     &   0.018   &   $144^2 \times 288$   &   $3.0 \times 10^1$\\
2   &   0.5     &   0.05   &   $144^2 \times 288$   &   $5.4 \times 10^1$\\
3   &   0.7     &   0.10   &   $144^2 \times 288$   &   $10^2$\\
4   &   0.9     &   0.16   &   $144^2 \times 288$   &   $1.7 \times 10^2$\\
\cmidrule(lr){1-4}
\rowcolor{Gray}
     &   \multicolumn{3}{c}{Constant bubble size, $d_b=0.5$ mm}   &  \cellcolor{White} $3.0 \times 10^2$\\
No.    &   $L_x=L_y$ [mm]  &  $\Theta$    &   $N_x \times N_y \times N_z$ &   $5.4 \times 10^2$\\
\cmidrule(lr){1-4}
5   &   3     &   0.021    &   $216^2 \times 288$   &   $1.0 \times 10^3$\\
6   &   2     &   0.0.05    &   $144^2 \times 288$   &   $1.7 \times 10^3$\\
7   &   1.33     &   0.11    &   $96^2 \times 288$   &   $3.0 \times 10^3$\\
8   &   0.89     &   0.25    &   $64^2 \times 288$   &   $5.4 \times 10^3$\\
9   &   0.7     &   0.40     &   $48^2 \times 288$   &   $10^4$\\
10  &   0.5925     &   0.60   &   $42^2 \times 288$   &     \\

 \end{tabular}
 \caption{Simulation parameters for cases with  varying bubble departure diameter $d_b$ at constant bubble spacing, and with varying bubble spacing $S=L_x=L_y$ at a fixed bubble departure diameter. The domain height is $L_z=4$ mm for all the simulation cases. At each configuration, the simulations are performed at 13 different current densities as listed in the last column, leading to 130 simulation cases in total.}
  \label{Table1_ch2}
 \end{table}

	\subsection{Governing equations}
	\subsubsection{Carrier phase}\label{sec:eq_carrier}
	\noindent Three-dimensional transient incompressible Navier-Stokes equations in Cartesian coordinate system are adopted to solve for the velocity field, $\mathbf{u}$, which include the momentum equation

	\begin{align}\label{eq:NS}
		\frac{\partial \mathbf u}{\partial t} +   \boldsymbol\nabla \cdot \left(\mathbf{uu} \right) = - \boldsymbol\nabla p +  \nu  \boldsymbol\nabla^2 \mathbf{u} + \mathbf{f_u},
	\end{align}
    \noindent and the continuity equation 
	\begin{align}\label{eq:continuty}
		\boldsymbol	\nabla \cdot \mathbf{u} = 0. 
	\end{align}
	
	\noindent Here, $ \boldsymbol \nabla$ is the gradient operator vector, $p$ is the modified kinematic pressure (i.e., the total pressure with the hydrostatic pressure subtracted), $\nu$ is the kinematic viscosity of the solution and $\mathbf{f_u}$ denotes the IBM direct forcing term used to enforce the velocity boundary conditions on the bubble interface. 
	
	In the most general case, the distribution of the $\acid$ is obtained by solving the advection-diffusion-migration equation for its constituent ions ($\hion$, $\sulphate$). Yet, for a binary electolyte it is possible to simplify the problem by assuming electroneutrality throughout the electrolyte \citep{Dickinson2011}, thus eliminating the migration terms between the ions transport equations. Hence, a single transport equation for $\acid$ with an effective diffusivity is obtained \citep{Sepahi2022,Morris1963}. Additionally accounting for $\hyd$, the transport of each substance, $C_j$, in the system can be described by an effective advection-diffusion equation as
	
	\begin{align}\label{eq:ADM}
		\frac{\partial C_j}{\partial t}  + \boldsymbol\nabla \cdot \left( \mathbf u C_j\right)= D_j\boldsymbol \nabla^2 C_j + \mathbf{f}_{C_j}, 
	\end{align}
	
	\noindent where the subscript $j=(s, \mathrm{H_2})$ refers to $\acid$ and $\hyd$, respectively.  Here, $\mathbf{f}_{C_j} $ is the IBM forcing term to enforce the respective gas-liquid interfacial condition for each substance and will be explained in $\S$ \ref{sec:eq_dispersed}. The effective diffusivity of $\acid$ can be obtained from the mass diffusivities, $D_k$, and ionic valences, $z_k$, of the ions  ($k=1,2$ denotes $\hion$ and $\sulphate$, see table \ref{Table2_Ch2} for ions diffusivity) as
	
	\begin{align}\label{eq:effective_diffusity}
		D_s=\frac{D_1 D_2(z_1 - z_2)}{z_1D_1 -z_2D_2}.
	\end{align} 

\tabcolsep 15 pt
\begin{table}
\begin{tabular}{llll}
			Symbol  & Description    &   Value    & Unit\\
			\hline
			$C_{\mathrm{s,0}}$    &  $\acid$ initial concentration   &   500 & $\mathrm{mol/m^{3}}$\\
   			$C_{\hyd,0}$    &  $\hyd$ initial concentration &  0 & $\mathrm{mol/m^{3}}$\\
         	$\Csath$    &  $\hyd$ saturation concentration &  0.72 & $\mathrm{mol/m^{3}}$\\
			$T_0$    &   Ambient temperature &   298 &   K \\
			$P_0$    &   Ambient pressure    &   1   &   $\mathrm{bar}$ \\ 
   			$\mathcal R$    &   Gas universal constant    &   8.314   &   $\mathrm{J/mol \cdot K}$ \\ 
			$\rho_L$ &    Electrolyte density    &   1030    &    $\mathrm{kg/m^{3}}$ \\ 
   			$\rho_G$  &  $\hyd$ density (for simulations)   &   1   &   $\mathrm{kg/m^{3}}$ \\ 
      		$\mu$   &   Electrolyte dynamic viscosity   &    $1.03 \times10^{-3}$    &  $\mathrm{kg/s \cdot m}$ \\ 
      		$\nu$   &   Electrolyte kinematic viscosity   &    $1.0 \times10^{-6}$    &  $\mathrm{m^2/s}$ \\ 
			$D_{\mathrm{H^+}}$    &   $\mathrm{H^+}$ diffusivity  &    $9.308 \times10^{-9}$  &  $\mathrm{m^2/s}$ \\ 
			$D_{\sulphate}$    &   $\sulphate$ diffusivity  &    $1.0 \times10^{-9}$  &  $\mathrm{m^2/s}$ \\ 
			$D_{\mathrm{s}}$    &   $\acid$ diffusivity  &    $2.47 \times10^{-9}$  &  $\mathrm{m^2/s}$ \\ 
			$D_{\hyd}$    &   $\hyd$ diffusivity  &    $3.7 \times10^{-9}$  &  $\mathrm{m^2/s}$ \\ 
			$k_{h,\hyd}$  &   $\hyd$ Henry's constant &   $7.2 \times 10^{-6}$ &    $\mathrm{mol/m^{3} \cdot Pa}$\\
\end{tabular}
   \caption{Physical properties of the analyzed system.}
   \label{Table2_Ch2}
\end{table}
 
	The no-slip impermeable condition is applied on the electrode. A uniform current density, $i=I/A_e$, where $I$ and $A_e$ are respectively the overall electric current and electrode surface area, is spread on the electrode surface except for an inactive area with instantaneous radius of $R_a=0.75R$ \citep{Vogt2015} underneath the bubble where zero current density is applied ($R$ is the bubble radius as a function of time, see figure \ref{fig:Numerical_sketch}($b$)). The current density in the outer region is therefore corrected slightly as the bubble grows in order to keep the overall electric current $I$ constant throughout the simulations. The cathodic set of boundary conditions for $C_j$ reads \citep{Sepahi2022,Morris1963}
	
	\begin{align}\label{eq:CathodBC}
		-\frac{i}{(n_e/s_1)F}=2D_1\left(1-\frac{z_1}{z_2}\right)\left(\frac{\partial C_s}{\partial z}\right)_{z=0},\\
		\frac{i}{(n_e/s_{\mathrm{H_2})}F}=D_{\mathrm{H_2}}\left(\frac{\partial C_{\mathrm{H_2}}}{\partial z}\right)_{z=0}.
	\end{align}
	
	\noindent Here, $n_e=2$ is the number of the transferred electrons in the cathodic reaction \eqref{eq:HER}, $s_1=2$ and $s_\hyd=1$ are the stoichiometric coefficients of the ions and $F=96485 ~\mathrm{C mol^{-1}}$ is the Faraday constant. After simplification, the corresponding cathodic flux $J_j=-D_j\left( \frac{\partial C_j}{\partial z} \right)_{z=0}$ for each species can be related to the current density via the Faraday constant as
	
	\begin{align}\label{eq:CathodFlux}
		J_{s}=\frac{1}{3} \frac{i}{F}\frac{D_s}{D_1},  ~\quad \text{and} \quad  J_{\mathrm{H_2}}=-\frac{i}{2F}.
	\end{align}

	While generally the boundary conditions at the top boundary are free-slip no-penetration and constant concentrations for the velocity and scalar fields, respectively, a remedy is required to allow the bubble pass the top boundary. For this purpose, we momentarily change the boundary condition to an in-outflow condition once the bubble arrives at the top boundary and revert back to the original boundary conditions once the bubble has left the computational box. The bubble passes through the top boundary with a constant velocity equal to its rise velocity before the boundary condition switch. We ensured that the computational domain is sufficiently high such that this procedure has negligible influence on mass transfer processes at the electrode. Moreover, periodic boundary conditions for the velocity and concentration fields are employed in the lateral directions of the computational domain. The choice of these boundary conditions is such that the corresponding pure-diffusion problem reaches a steady state for which an analytical self-similar solution exists \citep{Carslaw1959,VanDerLinde2017}. Thus, the known mass transfer rate of the pure-diffusion problem can be served as a base system for comparison of mass transfer change resulting from the bubbly flows within the electrolyte (see $\S$ \ref{sec:electrode}). 
	
	In order to numerically obtain the solution of \eqref{eq:continuty}, \eqref{eq:NS}, and \eqref{eq:ADM}, a second-order accurate central finite-difference scheme is employed for spatial discretization and time-marching is performed with a fractional step third-order accurate Runge-Kutta scheme \citep{Kim1985,Verzicco1996}. A multiple-resolution strategy \citep{Ostilla-Monico2015}, with refinement factor of two for the scalar fields, is used to solve the momentum and scalar equations, to cope with the fact that the mass diffusivity is several orders of magnitudes smaller than the momentum diffusivity. The grid is equally spaced in all directions. A grid independence check has also been performed and is reported in Appendix \ref{sec:grid_check}.

	\subsubsection{Dispersed phase}\label{sec:eq_dispersed}
	\noindent Numerically, we represent the growth and rise phases of the bubbles but circumvent the intricacies of the nucleation process by initializing the bubbles with a finite size of $R_0=50~\mu m$. During the growth phase, the expansion rate of the bubble is directly related to the diffusive transport of the dissolved gas across the gas-liquid interface which is determined by Fick's law.  Balancing the rate of the change of mass within the bubble and the diffusive flux of hydrogen across the interface as
	\begin{align}\label{eq:bub_flux}
		\dot{N}_b =\frac{P_0}{\mathcal{R} T_0 } 4\pi R^2 \frac{dR}{dt}=\int_{\partial V} D_{\mathrm{H_2}} \boldsymbol \nabla C_{\mathrm{H_2}} \cdot \hat{\mathbf{n}}_b \, \dd A,
	\end{align}
	yields the bubble growth rate
	\begin{align}\label{eq:Bubble_growth}
		\frac{\dd R}{\dd t}=\frac{\mathcal R T_0}{P_0} \frac{1}{4\pi R^2}\int_{\partial V} D_{\mathrm{H_2}} \boldsymbol \nabla C_{\mathrm{H_2}} \cdot \hat{\mathbf{n}}_b \, \dd A,
	\end{align}
	\noindent where $\mathcal R$, $T_0$ and $P_0$ are the gas universal constant, ambient temperature and pressure respectively. $R$ is the instantaneous radius of the bubble and $ \hat{\mathbf{n}}_b$ is the unit normal vector at the surface $\partial V$ of the bubble. Here, we assume a constant pressure inside the bubble throughout the growth phase, which is valid since for the range of bubble sizes $R \geq 50~ \mu$m the Laplace pressure is negligible compared to the ambient pressure of $P_0 = 1$ bar. We further neglected inertial effects on the pressure inside the bubble. This is confirmed  to be appropriate by computing the inertial terms of Rayleigh-Plesset equation $\rho_L(R\dot R +3\dot R ^2/2)$ \citep{Prosperetti1982}. For the largest bubble growth rates encountered in our simulations the corresponding change in the bubble pressure does not exceed 0.2 Pa, which is small compared to $P_0$. 
	
	The bubble detaches and rises under the influence of buoyancy in the electrolyte solution after growing to a prescribed departure diameter, $d_b$. Note that we do not consider a potential bubble growth during the rise phase. Given the short rise times ($\sim 0.1$ s) compared to the residence time of the bubble on the electrode ($\sim 1-100$ s) and the significantly lower hydrogen concentrations outside the boundary layer  at the electrode, this is has hardly any effect on our results. The bubble is treated as a spherical rigid particle during the rising phase and its deformation is disregarded owing to its small size ($d_b < 1~ \mathrm{mm}$), i.e. surface tension forces, which maintain the spherical form of the bubble, are predominant over inertia and drag forces in the ascent (Weber and Capillary numbers are significantly lower than unity). We solve for the translational velocity of the bubble, $\mathbf u_b$, which we assume to be governed by the Newton's second law of motion as 
	
	\begin{align}\label{eq:NewtonBubble}
		\rho_g V_b \frac{\dd \mathbf u_b}{\dd t} +\mathbf F_v=\int_{\partial V_b} \boldsymbol \tau \cdot \hat{\mathbf{n}}_b \, \dd A + \left(\rho_G - \rho_L\right)V_b\mathbf{g} + \mathbf F_v,
	\end{align}
	
	\noindent where 
	
	\begin{align}\label{eq:ub_param}
		\mathbf u_b=\frac{\dd \mathbf x_b}{\dd t}, \quad \boldsymbol \tau = -p \mathbf{I} + \mu \left(   \boldsymbol \nabla \mathbf{u} + \boldsymbol \nabla \mathbf{u}^T\right) \quad \text{and} ~   \mathbf F_v=C_v \rho_L \frac{\dd \mathbf u_b}{\dd t} .
	\end{align}
	
	\noindent Here, $\mathbf x_b$ is the bubble centroid position, $\rho_G$ and $\rho_L$ are the gas and fluid densities, respectively, $V_b$ is the bubble volume after detachment, and $\boldsymbol \tau$ is the stress tensor for Newtonian fluids. Following \citet{Schwarz2015}, the virtual mass term, $\mathbf F_v$ with $C_v > 0$ is added to both sides of (\ref{eq:ub_param}). Note that $F_v$ is evaluated explicitly on the right hand side but implicitly on the left hand side, thus increasing the diagonal dominance of the inversion coefficient in the case of very low gas-fluid density ratio $\Gamma=\rho_G/\rho_L \ll 1$. In the present work, the latter is set to $\Gamma=0.001$ and the bubble motion equation is solved with $C_v=0.5$. We resort to virtual mass method with standard IBM here due to the rather simple wake flow of the light rising bubbles at low Reynolds number. A comparison between our simulation results and those of \citet{Schwarz2015}, presented in Appendix \ref{sec:virtual_mass}, demonstrates reasonable accuracy and reliability of this method for the problem under investigation in this study. However, in case of higher Reynolds in which wake instabilities lead to complex flow motion, one may consider using more robust but computationally much more demanding methods like IBM with strong coupling of fluid-structure interaction \citep{Borazjani2008} or IBM projection method \citep{Lacis2016,Assen2023}. It is worth noting that the hydrodynamic force on the bubble is related to the IBM forcing term, $\mathbf {f_u}$, as follows \citep{Uhlmann2005,Kempe2012,Breugem2012}
	
	\begin{align}\label{eq:HydrodynamicForces}
		\int_{\partial V_b} \boldsymbol \tau \cdot  \hat{\mathbf{n}}_b \, \dd A = -\rho_L \int_{V_b}\mathbf{f_u} \, \dd V + \rho_L \frac{\dd}{\dd t} \left(\int_{V_b} \mathbf u \, \dd V \right).
	\end{align}
	
	A set of the boundary conditions for the carrier phase on the bubble interface is required for the concentration and velocity fields. Saturation concentration based on Henry's law $\Csath=k_hP_0$, with $k_h$ being the Henry's constant for $\hyd$ and zero flux $\boldsymbol \nabla C_s \cdot  \hat{\mathbf{n}}_b =0$ for $\acid$ are applied on the bubble interface. Assuming a fully contaminated bubble \citep{Takagi2011}, the no-slip no-penetration condition is employed on the bubble interface ($\mid \mathbf x - \mathbf x_b  \mid=R$) such that the velocity $\mathbf{u}\vert_{\partial V}$ of a point on the bubble surface is given by
	
	\begin{align}\label{eq:Bubble_VBC}
		\mathbf u\vert_{\partial V} = \mathbf u_b + \frac{\dd R}{\dd t} \hat{\mathbf{n}}_b.
	\end{align}
	
	\noindent This relation is coupled to the mass transfer via \eqref{eq:Bubble_growth} to determine the bubble growth rate $\dd R/\dd t$. To ensure continuity within the domain during the bubble growth, the continuity equation needs to be revised by adding a source term in the bubble interior according to  
	
	\begin{align}
		\boldsymbol \nabla \cdot \mathbf u = \phi \frac{3}{R}\frac{\dd R}{\dd t},
	\end{align}
	
	\noindent where $\phi$ is an indicator function which undergoes a smooth transition from 0 to 1, based on a cut-cell method \citep{Kempe2012} for the cells outside and inside the bubble, respectively. This amendment is necessary for modeling expanding/shrinking boundaries using an incompressible solver with IBM. The same approach has also been adopted in the literature for simulation of flows with evaporating droplets \citep{Lupo2019,Lupo2020}. The local velocity field is still entirely divergence free outside the bubble and the nonzero divergence inside the bubble is irrelevant to the flow physics outside due the boundary conditions enforced on the gas-liquid interface. To ensure the global conservation of the mass in the course of the bubble growth, a small but non-zero uniform vertical velocity is prescribed at the top boundary such that the outflow rate equals the expansion rate of the bubble, similar to the simulations of evaporating droplets in the wall-bounded turbulent flows using IBM \citep{Lupo2020}.
	
	The bubble interface is discretized using another triangulated Lagrangian grid as depicted in figure \ref{fig:Numerical_sketch}($b$). The IBM method here is based on moving least squares (MLS) approach to conduct the interpolation and distribution of the direct forcing terms between the Eulerian and Lagrangian grids \citep{book_Liu,Vanella2009,Spandan2017}. The enforcement of the Dirichlet and Neumann conditions on the interface for $\mathrm{H_2}$ and $\mathrm{H_2SO_4}$ is performed employing a ghost-cell based IBM to ensure the conservation of the species \citep{Lu2018}. To validate these procedures, we verified that mass conservation for the hydrogen distribution is fullfilled in our simulations (see Appendix \ref{sec:hyd_consv}).  
 
	\subsection{Response parameters}\label{sec:response_param}
    \noindent The most basic response parameters relate to the transport of $\hyd$ away and $\acid$ towards the electrode. Since the respective rates of production and consumption at the electrode, $J_{\mathrm{H_2}}$ and $J_s$, are constant in time, the effective transport is reflected in the difference between the surface averaged concentrations of hydrogen, $\Cgt$ and electrolyte, $\Cst$, at the electrode surface and their respective initial values in the bulk ($C_{\hyd,0}$ and $C_{\mathrm{s},0}$). We can normalize these differences using the (constant) fluxes $J_j$ and the bubble diameter $d_b$ as reference scales to yield the Sherwood numbers
 	\begin{align}\label{eq:Sh_electrode}
		\Shget=\frac{J_\hyd d_b}{D_\hyd \left( \Cgt - \Csath \right)}, \quad \text{and} ~  	\Shst=\frac{J_{\mathrm{s}}d_b}{D_{\mathrm{s}} \left(C_{\mathrm{s},0} - \Cst \right)}.
	\end{align}
    Here and in the following the tilde symbol is used to differentiate time-dependent response parameters from the corresponding averages over a bubble-period without tilde where necessary. By introducing the boundary layer thickness $\tilde \delta_j=D_j \Delta \tilde C_j/J_j$, this Sherwood number can equivalently be expressed as $\tilde{Sh}_j = d_b/\tilde \delta_j$. For pure diffusion, $\tilde \delta_j$ ultimately reaches the cell height irrespective of the current density such that the same steady-state value of $\tilde{Sh_j}$ would be obtained for all cases without the effect of the bubbles. 

    Analogously, we characterise the mass transfer of hydrogen into the bubble using the bubble Sherwood number
	\begin{align}\label{eq:Sh_bubble}
		\Shgbt=\frac{2\dot R R}{\frac{\mathcal R T_0}{P_0}D_\hyd \left(\Cgt-\Csath\right)},
	\end{align}
	which employs the instantaneous bubble diameter, $2R$, the surface area, $4\pi R^2$, and the concentration difference between the electrode and bubble interface, $\left(\Cgt - \Csath \right)$, for normalization of the mass flux into the bubble given by \eqref{eq:bub_flux}.

	A final important output is the fraction of the total hydrogen produced that ends up in gaseous form, i.e. gets desorbed into the bubble \citep{Vogt1984_v1,Vogt1984_v2,Vogt2011a,Vogt2011b}. Mathematically formulating this leads to an expression for the gas-evolution efficiency 
	\begin{align}\label{eq:fG}
		f_G=\frac{\frac{V_b}{\tau_c}}{\frac{\mathcal R T_0}{P_0}\frac{-i}{n_eF}A_e}=\frac{\dot V_G}{\frac{\mathcal R T_0}{P_0}\frac{-i}{n_eF}A_e},
	\end{align}
	where $\tau_c=\tau_g + \tau_r$ is the bubble lifetime, which comprises the bubble residence (growth) time, $\tau_g$, and the bubble rise time, $\tau_r$. $\dot V_G=V_b/\tau_c$ is the volumetric gas flux into the gas phase.

 	\begin{figure}
		\centering
		\includegraphics[width=1\columnwidth]{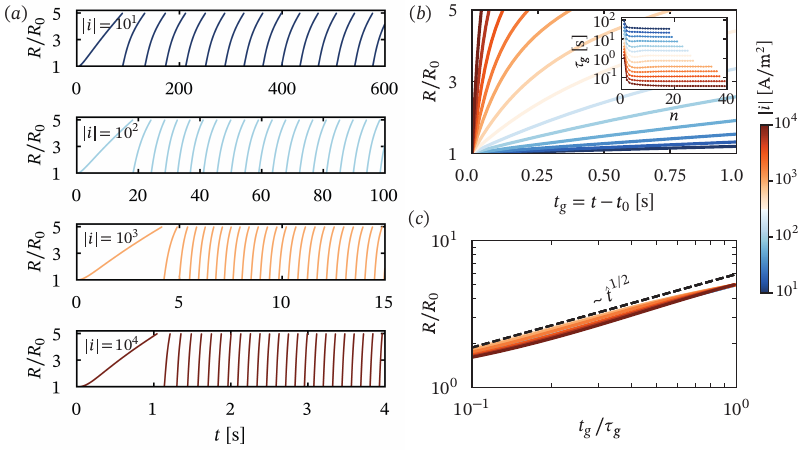}
		\caption{($a$) Radius of the successively growing bubbles as function of time for current densities $\lvert i \lvert =10^1,10^2,10^3$ and $10^4~\mathrm{A/m^2}$. The radius has been normalized with the initial size of the bubble used for the simulations, $R_0=50~\mu m$.  ($b$) Temporal evolution of the bubble radius at statistically steady-state for each current density in the range of $ 10^1<\lvert i \lvert<10^4 ~\mathrm{A/m^2}$. The magnitude of the current density is illustrated with the colormap.  $t_0$ is the start of the bubble life-time in each case and hence $t_g=t-t_0$ is the bubble age. The inset shows the bubble growth time, $\tau_g$, for the $n^{th}$ bubble.  ($c$) double-logarithmic plot of the bubble evolution curve for all the current densities. Time axis has been normalized with the growth time in the steady state as shown in the inset of panel ($b$).}
		\label{fig:radius}
	\end{figure}

 	\begin{figure}
		\centering
		\includegraphics[width=1\columnwidth]{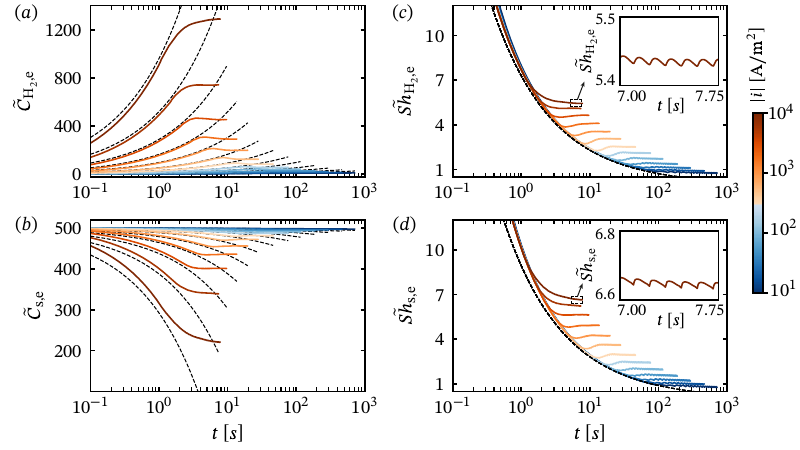}
		\caption{Temporal evolution of hydrogen ($a$) and electrolyte ($b$) averaged concentrations at the electrode surface for bubble departure diameter of $d_b=0.5$ mm and spacing of $S=2$ mm for all the investigated current densities. Broken black lines represent the solution of  the pure diffusion problem in a semi-infinite medium with constant flux condition at the boundary, calculated using \eqref{eq:semi_infinite}. Corresponding Sherwood numbers of simulations and pure-diffusion problem for hydrogen ($c$) and electrolyte ($d$) transport computed based on \eqref{eq:Sh_electrode}. Insets in panels ($c$) and ($d$) show a closer view of Sherwood variation for the highest current density in the statistically steady state. Current density at each case is distinguished using the colormap whose range is shown in the colorbar.}
		\label{fig:C_Sh}
	\end{figure}

	\begin{figure}
		\centering
		\includegraphics[width=1\columnwidth]{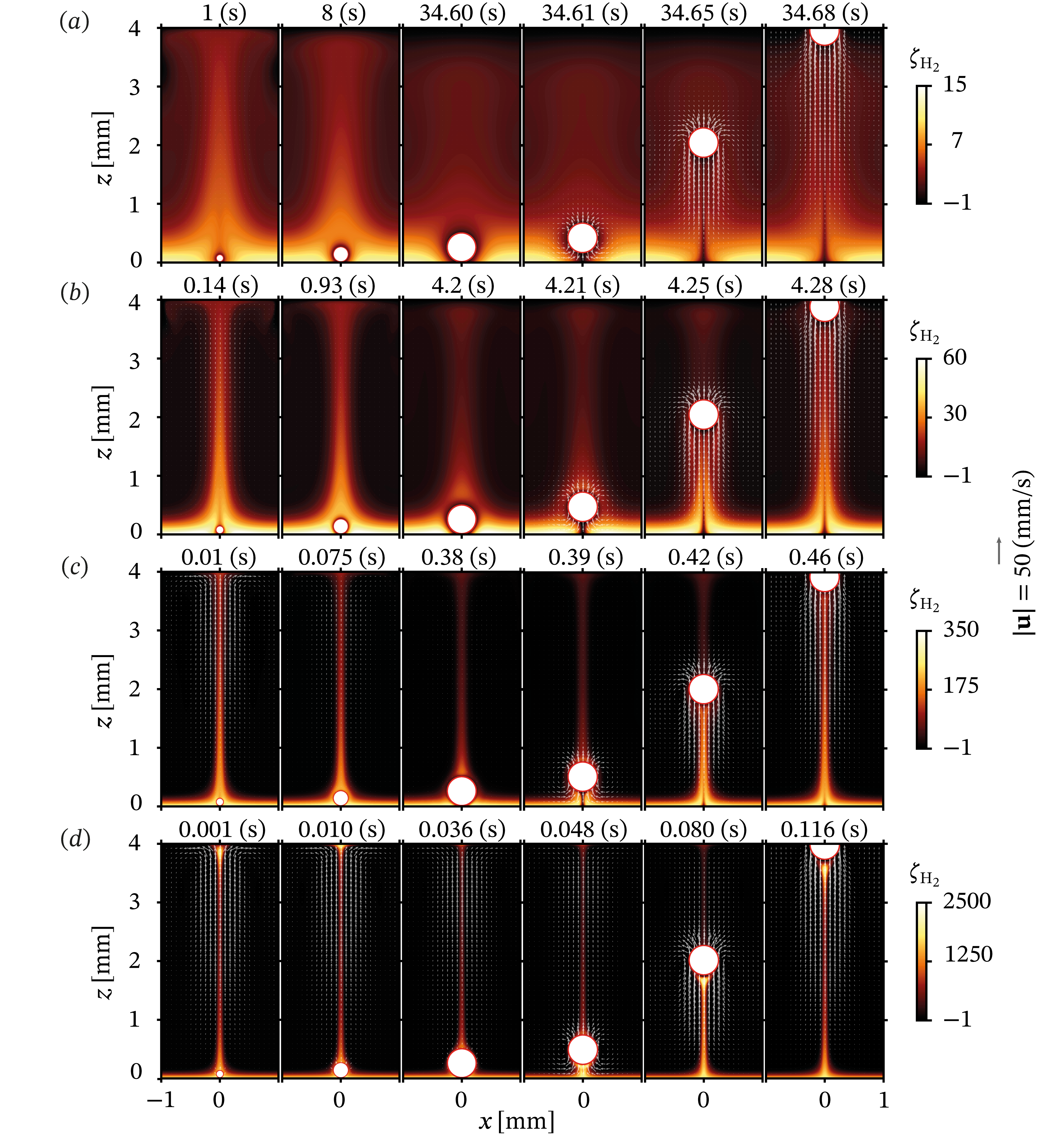}
		\caption{Snapshots of the hydrogen and velocity distributions in the equilibrium state at different stages of the bubble life-time for current densities of ($a$) $10^1$, ($b$) $10^2$, ($c$) $10^3$ and ($d$) $10^4~\mathrm{A/m^2}$. Bubble break-off diameter is $d_b=0.5$ mm and spacing is set at $S=2$ mm. In all cases, the first three panels cover the bubble growth and the last three the bubble rise time. The supersaturation level, $\zeta_{\mathrm{H_2}}$, is shown using the colorbar. The superimposed vectors represent the induced velocity field by the growth and rise of the bubbles in the electrolyte. The velocity scale provided at the right of the figure applies to all panels.   }
		\label{fig:Gas_profile}
	\end{figure}
 
 	\begin{figure}
		\centering
		\includegraphics[width=1\columnwidth]{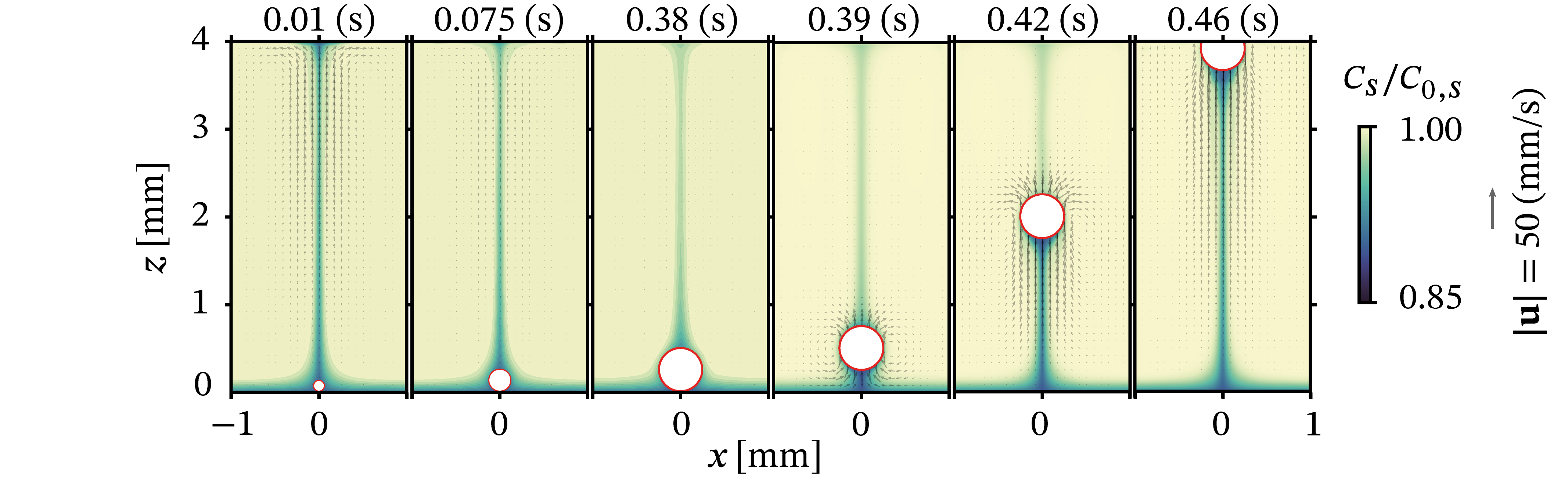}
		\caption{Snapshots of the electrolyte distribution for the case ($\vert i \vert =10^3~\mathrm{A/m^2}$) shown in figure \ref{fig:Gas_profile}($c$).  }
		\label{fig:Cs_proflie}
	\end{figure}

	\begin{figure}
		\centering
		\includegraphics[width=1\columnwidth]{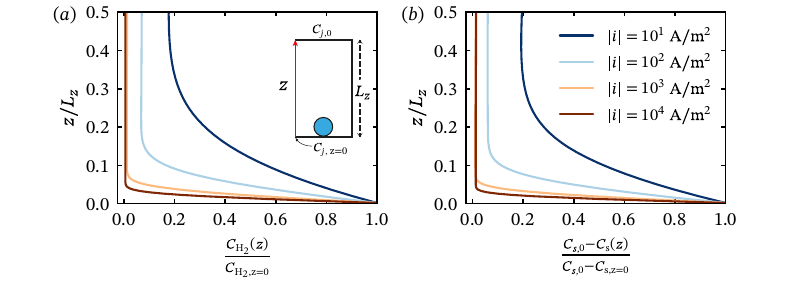}
		\caption{Vertical profiles of normalized hydrogen ($a$) and electrolyte ($b$) concentration half-way between adjacent bubbles (see the sketch in panel ($a$)) at the instant of bubble break-off. The profiles are captured at statistically steady state for different current densities. }
		\label{fig:BL}
	\end{figure}

	\section{Bubble dynamics and mass transfer at the electrode}\label{sec:electrode}
	\noindent To begin with, we present the simulation results for a bubble departure diameter of $d_b=0.5$ mm and spacing $S=2$ mm. The physical properties of the system are set in accordance to table \ref{Table2_Ch2}. Figure \ref{fig:radius}($a$) shows the growth dynamics of successively generated bubbles on the electrode at four different current densities. At each current density, the first few bubbles show a slower growth while the supersaturation level of the gas in the electrode boundary layer is building up and the growth pattern becomes more repetitive at later times. This is also reflected in the bubble growth time, which drops initially, but remains constant for subsequent bubbles later on (see inset of figure \ref{fig:radius}($b$)). These observations are indicative of an equilibrium state, in which the time-averaged mass transport and gas production rates at the electrode surface are balanced, leading to the repetition of the same growth dynamics for bubbles evolving in sequence. The bubble size evolution at statistically steady-state is plotted and compared in figure \ref{fig:radius}($b$) for different current densities. These curves have been taken at times when the bubble residence time, $\tau_g$, no longer varies with bubble number $n$, as depicted in the inset. Despite the fact that the bubble growth time varies several order of magnitudes from 100 s to less than 0.1 s when increasing the current density from $10^1$ to $10^4~\mathrm{A/m^2}$, the growth dynamic pertaining to diffusion-limited growth, i.e. $R~\propto t^{1/2}$, is maintained \citep{Epstein1950,Scriven1959}. This is evidenced by the double-logarithmic plot of the bubble size evolution in figure \ref{fig:radius}($c$), where the time axis is normalized with $\tau_g$. In this form, all cases approximately collapse onto a single curve that is in good agreement with the $1/2$ power law. 

	Next, we look into the mass transfer rate at the electrode by tracking the spatially averaged concentrations on the electrode surface in time, as shown in figures \ref{fig:C_Sh}($a$) and \ref{fig:C_Sh}($b$) for $\hyd$ and $\acid$, respectively. As the reaction proceeds, the hydrogen concentration increases in time in contrast to the electrolyte concentration, which is depleted at the electrode. For the one-dimensional pure diffusion problem in the absence of the bubbles (diffusion in a semi-infinite medium with constant flux on the boundary) the analytical solution gives the time evolution of the cathodic concentrations, $\tilde{C}^{\ast}_{j,\mathrm{e}}$, as \citep{Bejan1993}
	\begin{align}\label{eq:semi_infinite}
		\tilde{C}^{*}_{j,\mathrm e}(t)-C_{j,0}=2J_j \sqrt{\frac{t}{\pi D_j}},
	\end{align}
	which has been provided for comparison at each current density in the figure \ref{fig:C_Sh}($a$) and \ref{fig:C_Sh}($b$). Small differences between this solution and the simulation results are related to the presence of the adhering bubble on the electrode and the inactive area underneath it, which alters the local concentrations slightly. Major deviations from the analytical solution occur after the departure of the first bubble, which leads to significantly enhanced mixing. As a result, fresh electrolyte is transported to the electrode, replacing the gas-enriched and electrolyte-depleted solution there. Eventually, the system reaches an equilibrium in which the reaction and transport rates are balanced, such that the cycle-averaged concentrations remain constant in time. 
	
	A comparison of the behaviour for different current densities $i$ is best done using the transient Sherwood numbers \eqref{eq:Sh_electrode} plotted in figures \ref{fig:C_Sh}($c$) and \ref{fig:C_Sh}($d$) for $\hyd$ and $\acid$, respectively. Prior to the first bubble departure form the electrode surface, time-dependent Sherwood numbers collapse to a single curve regardless of the current density, as do those pertaining to the analytical solution of the pure diffusion problem. The bifurcation from the main trend happens after the detachment of the first bubble, i.e. transition to the convection, which takes place earlier at higher current density due to the higher oversaturation of the dissolved gas in the electrode boundary layer and faster bubble growth. Once the system is at equilibrium and the bubble generation rate no longer changes, the Sherwood numbers also approach an equilibrium value. Small oscillations around this value occur within each bubble cycle (see insets for the highest current density). For these, the minimums of $\Shjt$ correspond to the detachment times after which the Sherwood numbers immediately increase and the maximums are the instants when the bubble lifetime starts, followed by a slow decrease during the growth time. Furthermore, due to the higher frequency of bubble generation and hence stronger mixing in the electrolyte, the effective mass transfer rate at the electrode, reflected in the values of $\Shjt$ in equilibrium, is significantly enhanced at higher current densities.

	In order to provide insight into flow structure and  scalar distribution in the equilibrium state, figure \ref{fig:Gas_profile} displays snapshots of the  hydrogen supersaturation, $\zeta_{\mathrm{H_2}}=C_{\hyd}/\Csath-1$, overlaid with velocity vectors at different stages of the bubble evolution and for varying current densities. For the case with $\vert i \vert=10^3~\mathrm{A/m^2}$, corresponding plots for the electrolyte concentration distribution are provided in figure \ref{fig:Cs_proflie}. At this current density, a maximum electrolyte depletion of $\approx15\%$ occurs at the electrode and even in the most extreme case with $\vert i \vert=10^4~\mathrm{A/m^2}$, this  value does not exceed $\approx 70\%$, meaning that the electrolyte concentration remains finite in all cases even though the diffusion limited current density, $\vert i \vert_{\mathrm{diff}} = n_e F D_s C_{s,0}/H = 59.7~\mathrm{A/m^2}$, is exceeded significantly. The associated transport enhancement is due to a large-scale convective pattern that is  established during the rise stage, with an up-draught stream in bubble column, downwelling flow along the (periodic) side-walls, and wall-parallel flow close to the electrode. At low current density (figure \ref{fig:Gas_profile}($a$)), the bubble driving is highly intermittent as the convective motion dissipates during the long growth period. However, as the latter becomes shorter for larger $i$ (figures \ref{fig:Gas_profile}($b$) and \ref{fig:Gas_profile}($c$)), the flow becomes more and more continuous and a strong circulation is visible throughout the entire bubble cycle at $\vert i \vert = 10^4~\mathrm{A/m^2}$ in figure \ref{fig:Gas_profile}($d$). The convective pattern counteracts the penetration of the electrode boundary layer into the bulk by advecting the fresh electrolyte towards the electrode. This effect is stronger at higher currents due to the higher frequency of bubble formation driving a stronger flow. This can be also appreciated from figure \ref{fig:BL}($a$) and \ref{fig:BL}($b$), which compares the vertical profiles of normalized $\hyd$ and $\acid$ at location halfway between adjacent bubbles, where an appreciable drop in the electrode boundary layer thickness with increasing current density is observed, consistent with an enhanced mass transport.

   	\begin{figure}
		\centering
		\includegraphics[width=1\columnwidth]{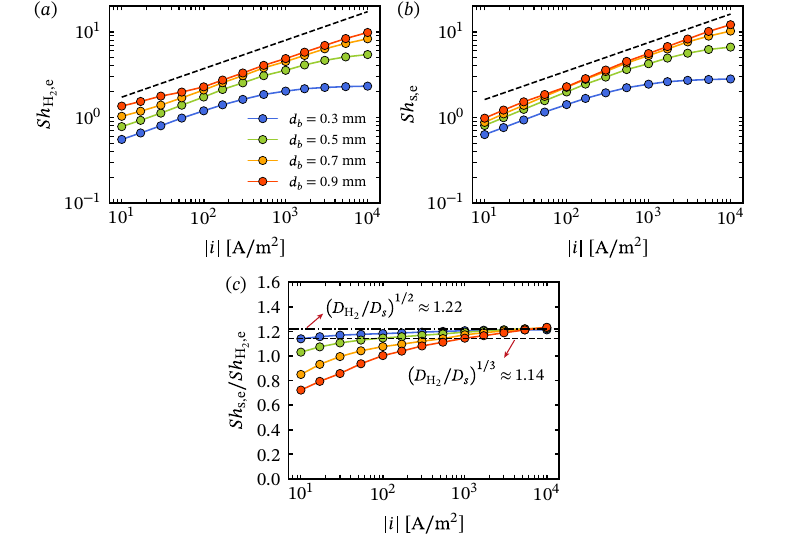}
		\caption{Sherwood number of ($a$) hydrogen and ($b$) electrolyte transport averaged over an entire bubble lifetime in the statistically steady state, as a function of current density for different bubble break-off diameter, $d_b$. The broken lines indicate the power law relation $Sh_j \sim i^{1/3}$ for reference. ($c$) Ratio of electrolyte to hydrogen Sherwood numbers versus the current density at different bubble diameters. Dashed and dashed-dotted lines correspond to $(D_\hyd/D_s)^{1/3}$ and $(D_\hyd/D_s)^{1/2}$, respectively, for comparison.}
		\label{fig:Sh_db}
	\end{figure}

	\subsection{Current dependence of the Sherwood number and bubble size effect}\label{sec:Bub_size}
 
\begin{figure}
		\centering
		\includegraphics[width=1\columnwidth]{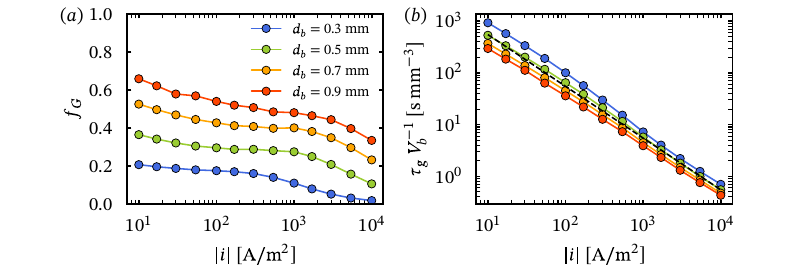}
		\caption{ ($a$) Gas-evolution efficiency, $f_G$ as a function of current density for different bubble break-off diameter, $d_b$. ($b$) Bubble residence time, $\tau_g$, compensated with bubble departure volume, $V_b$, as a function of current density for different values of $d_b$. The broken line indicates the power law of $\tau_g \sim i^{-1}$. }
		\label{fig:tau_G_db}	
	\end{figure}

 	\noindent Next, we consider the current dependence of the Sherwood numbers of hydrogen and electrolyte transport, averaged over an entire bubble lifetime in the statistically steady state, which are plotted in figure \ref{fig:Sh_db}($a,b$), respectively. Apart from the case with $d_b = 0.5$ mm considered so far, these figures also include results for other bubble departure diameters. The trend of increasing $Sh_j$ with increasing $i$, which was already evident in figures \ref{fig:C_Sh}($c,d$) for $d_b = 0.5$ mm, is consistently observed for all these cases. The current dependence approximates a power law scaling of $Sh_j \sim i^{1/3}$ especially for larger bubbles, but deviations occur for smaller bubbles at high current densities, where $Sh_j$ increases significantly slower. It is further interesting to examine how $\Shge$ and $\Shs$ relate to each other, which we do by plotting the ratio $\Shs/\Shge$ in figure \ref{fig:Sh_db}($c$). Given that $\Shs/\Shge = \delta_\hyd/\delta_s$,  one expects this ratio to yield a constant of either $\left( D_\hyd/D_s\right)^{1/2}$ (for diffusive transport) or $\left( D_\hyd/D_s\right)^{1/3}$ (for convection given that the Schmidt number $Sc_j = D_j/\nu$ is large \citep{Bejan1993}) for a single-phase flow. In the present simulations, $D_\hyd/D_s =1.5$, such that the resulting values (1.22 and 1.14) do not differ significantly. In our results in figure \ref{fig:Sh_db}($c$), a ratio of comparable magnitude is attained for the smallest bubbles and similar values are also approached for the cases with larger $d_b$ at successively larger magnitudes of $i$. The deviation from the single-phase value is related to the fact that the electrolyte is only transported in solution while hydrogen is also carried inside the bubble. It is therefore most pronounced at low current densities and for large bubble sizes since for these cases the fraction of gas transported in the bubbles is largest as the plot of $f_G$ in figure \ref{fig:tau_G_db}($a$) confirms. The gas efficiency decreases significantly with decreasing bubble size, but is only a weak function of the current density especially for $i\lessapprox 10^3\mathrm{A/m^2}$. From gas-evolution efficiency relation, \eqref{eq:fG}, it is deduced that $\tau_g \sim V_b \left(f_G~i\right)^{-1}$, considering a constant rise time ($\tau_c$) for the bubbles with the same size.  Given the weak dependence of $f_G$ on $i$, the scaling of $\tau_g/V_b \sim i^{-1}$ holds reasonably well for all the cases shown here, as can be seen from figure \ref{fig:tau_G_db}($b$).
 
	\subsection{Effect of bubble spacing}\label{sec:Bub_space}
     \noindent Changing the bubble departure size, as was done in $\S$ \ref{sec:Bub_size}, has multiple effects since it affects bubble growth times and the flow, but also alters the effective bubble coverage $\Theta$. To disentangle these, we now fix the departure diameter of the bubble at $d_b=0.5$ mm and vary the box size $S$ to explore a range of $ 0.02 \le \Theta \le 0.56$. This resembles a change in the bubble population density, which in practice is tied to the current density and typically increases when $i$ is increased \citep{Vogt2005,Vogt2013}. Taking the advantage of the numerical simulations, we can explore the effect of this parameter independently here.

   	\begin{figure}
		\centering
		\includegraphics[width=0.975\columnwidth]{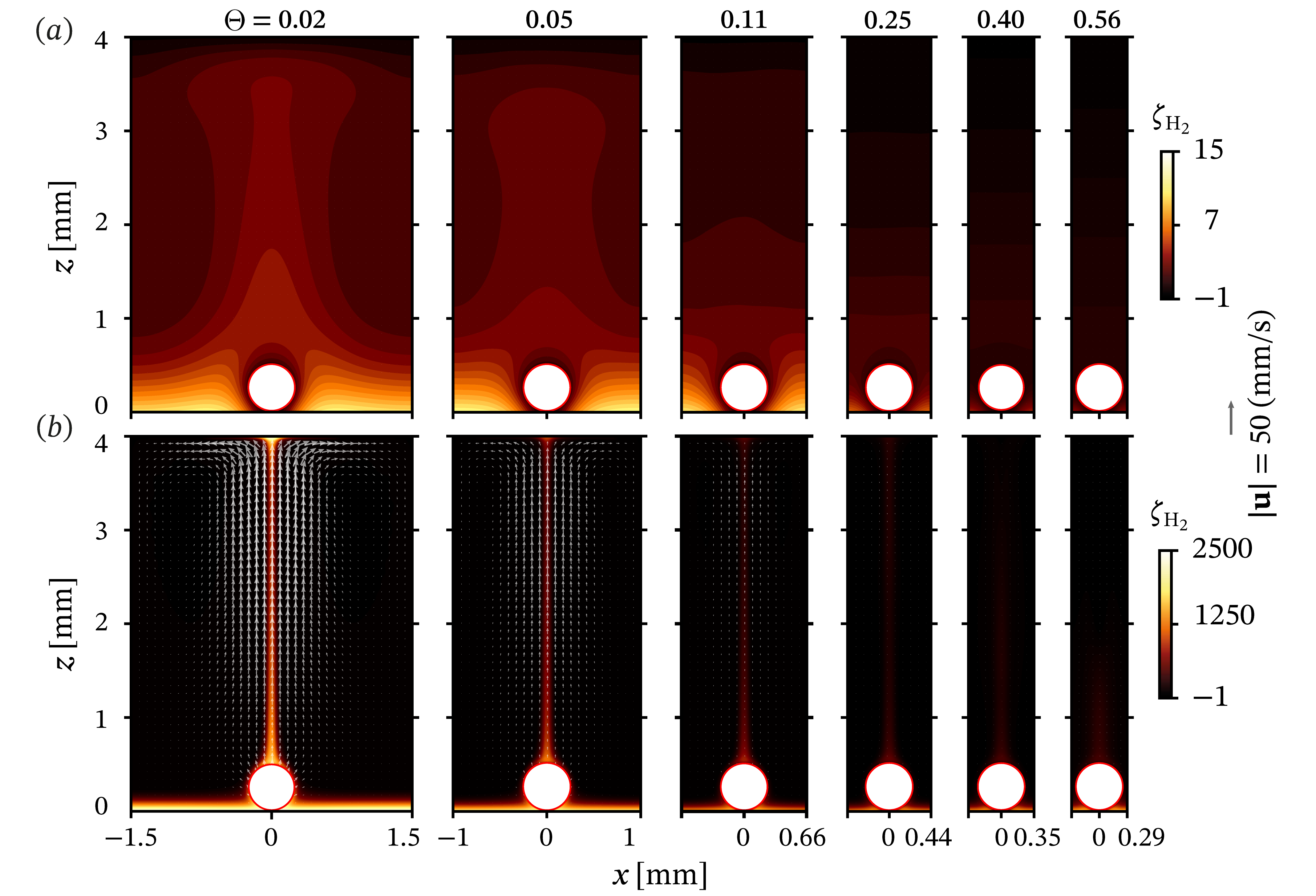}
		\caption{Snapshots of hydrogen supersaturation taken at the time of bubble detachment in the statistically steady state for $\vert i \vert=10^1$ ($a$) and $\vert i \vert=10^4~\mathrm{A/m^2}$ ($b$). The fractional bubble coverage is increased from left to right within the range $ 0.02 \le \Theta \le 0.56$ whose value is specified at top.  The velocity scale applies to all panels.}
		\label{fig:Gas_profile_Th}
	\end{figure}
 	\begin{figure}
		\centering
		\includegraphics[width=0.975\columnwidth]{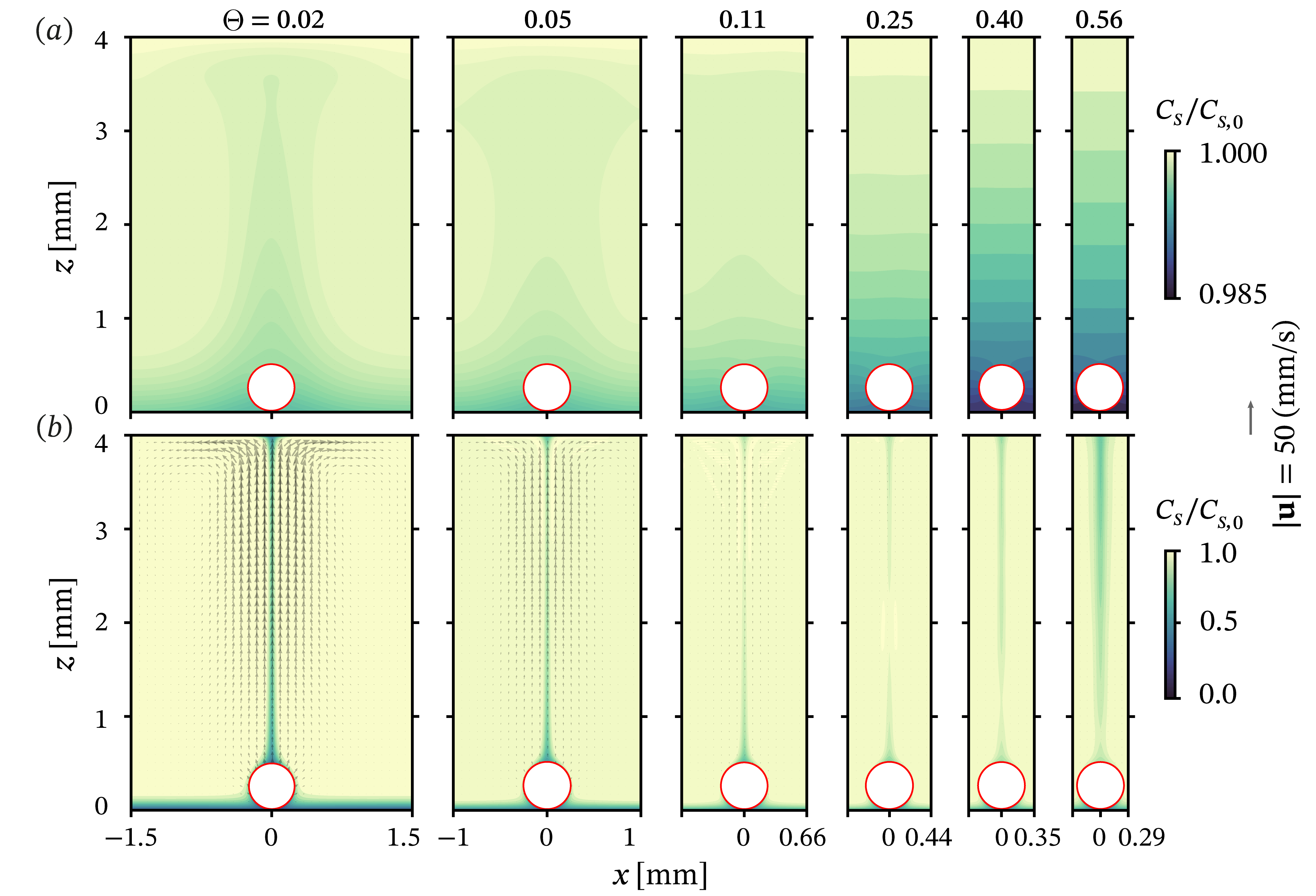}
		\caption{Snapshots of normalized $\acid$ distribution at the time of bubble detachment in the statistically steady state for $\vert i \vert=10^1$ ($a$) and $10^4\mathrm{A/m^2}$ ($b$). }
		\label{fig:Cs_profile_Th}
	\end{figure}
 
\begin{figure}
		\centering
		\includegraphics[width=1\columnwidth]{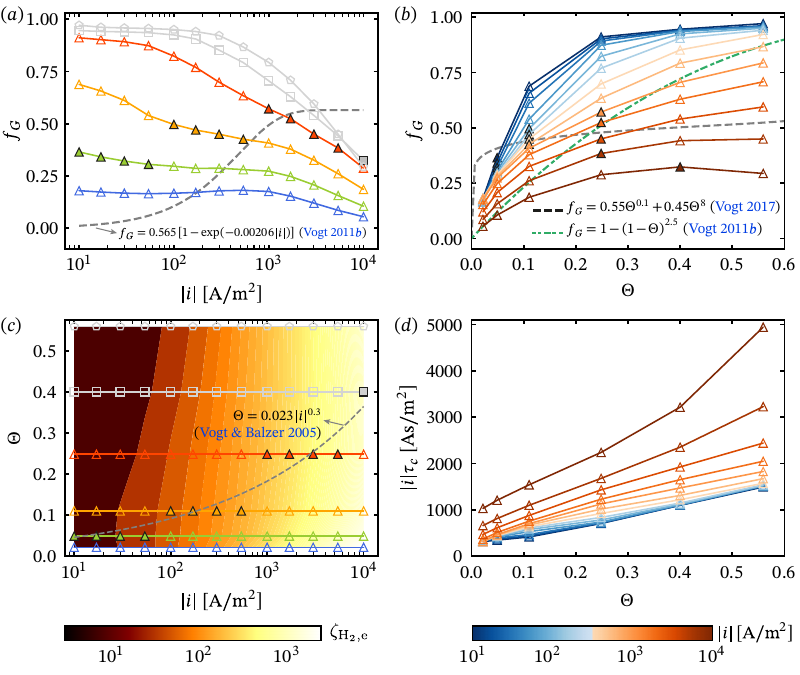}
		\caption{($a$) Gas-evolution efficiency, $f_G$, as a function of current density for varying bubble spacing (specified in terms of the fractional bubble coverage, $\Theta$). The bubble departure diameter has been fixed at $d_b=0.5$ mm. ($b$) Gas-evolution efficiency versus bubble coverage for varying current densities. ($c$) Hydrogen supersaturation on the electrode surface, $\zeta_{\mathrm{H_2,e}}$, for all the simulation cases with varying current density and bubble spacing. ($d$) Bubble lifetime, $\tau_c$, premultiplied with current density as a function of bubble coverage for varying current densities. The relevant empirical relations by Vogt $et~al.$ are provided with broken lines in the panels. The filled markers in panels ($a$) and ($b$) show the closest data to the empirical relation $\Theta=0.023 \vert i\vert^{0.3}$ \citep{Vogt2005} in panel ($c$), to highlight the more realistic cases.}
		\label{fig:fG_Theta}
	\end{figure}

    Figures \ref{fig:Gas_profile_Th} and \ref{fig:Cs_profile_Th} offer insight into how changing $\Theta$ affects the mass transport processes at the electrode by showing snapshots of the distributions of $\hyd$ and $\acid$, respectively, taken in the instant of bubble detachment after the system has reached a steady state. Figure \ref{fig:Gas_profile_Th}($a$) displays data for $\hyd$ at the lowest current density investigated ($\vert i \vert = 10^1~\mathrm{A/m^2}$). For this case, the boundary layers are thick due to the weak convective transport at low $\Theta$. However, as the bubble coverage is increased, the amount of dissolved hydrogen decreases and almost all the produced gas is contained in the bubble at $\Theta = 0.56$. This implies very efficient transport for $\hyd$ via the gas phase, but since the detachment frequency is low, the same does not hold for $\acid$ as can be seen from figure \ref{fig:Cs_profile_Th}($a$). Here, the depletion boundary layer is very thick with almost a linear gradient across the domain height. At the highest current density of $\vert i \vert = 10^4~\mathrm{A/m^2}$, the significantly shorter detachment period leads to a much stronger driving of the flow. Convective transport therefore prevails even at high $\Theta$, where $\tau_c$ tends to increase as the amount of hydrogen produced per bubble decreases for smaller bubble spacings (see figure \ref{fig:fG_Theta}($c$)). As a consequence, not only the hydrogen boundary layer (figure \ref{fig:Gas_profile_Th}($b$)) but also that for the electrolyte concentration (figure \ref{fig:Cs_profile_Th}) remain thin even at $\Theta = 0.56$.
 
	\begin{figure}
		\centering
		\includegraphics[width=1\columnwidth]{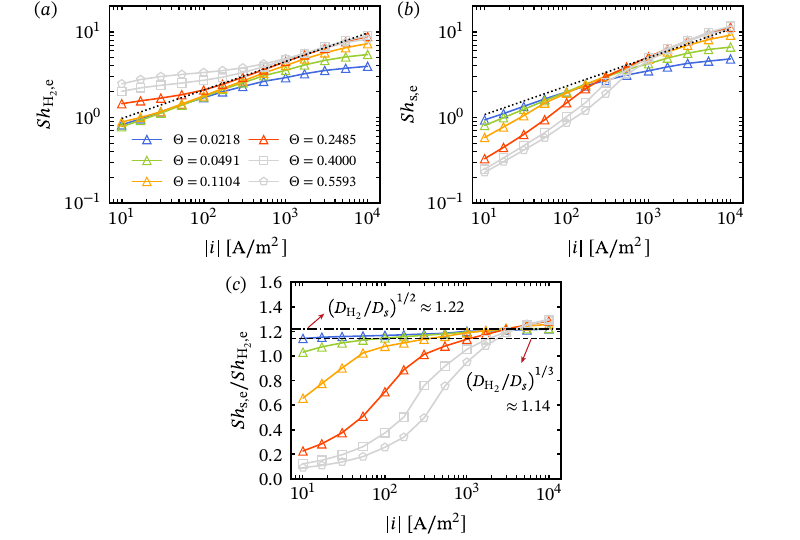}
		\caption{Sherwood number of ($a$) hydrogen and ($b$) electrolyte transport averaged over one bubble life-time in the statistically steady state, as a function of current density for different bubble spacings. The bubble departure diameter is fixed at $d_b=0.5$ mm and the range of fractional bubble coverage is $ 0.02 \le \Theta \le 0.56$ as specified in the legend.}
		\label{fig:Sh_Th}
	\end{figure}

    The trends observed in figures \ref{fig:Gas_profile_Th} and \ref{fig:Cs_profile_Th} are also reflected in the Sherwood numbers of $\hyd$ and $\acid$ plotted in figure \ref{fig:Sh_Th}($a$) and \ref{fig:Sh_Th}($b$). $\Shge$ increases with $\Theta$ throughout the whole range of current densities investigated. Again, the data generally approximate an $i^{1/3}$ scaling albeit with significant deviations at low $i$ and high $\Theta$ where the results significantly exceed this trend. Additionally, $\Shge$ falls below the $1/3$-scaling line for large current densities and low bubble coverage, which is in accordance with the trend observed in figure \ref{fig:Sh_db}($a$) for smaller $d_b$ for which the value of $\Theta$ is also reduced. For these higher currents, $\Shs$ behaves similar to $\Shge$ and this is also reflected in the ratio $\Shs/\Shge$ (figure \ref{fig:Sh_Th}($c$)) being close to those expected for single-phase transport. Interestingly, $\Shs/\Shge$ attains values even slightly larger than 1.22 for larger $\Theta$. Presumably, this is caused by the lower $\hyd$ concentration in the dissolved phase, which dominates the transport for these cases. Remarkably, the $\Theta$ trend of $\Shs$ at current densities $\vert i \vert\lessapprox 10^3~\mathrm{A/m^2}$ is opposite to that observed for the hydrogen transport in this regime with $\Shs$ decreasing for larger $\Theta$. The ratio $\Shs/\Shge$ drops to values as low as 0.1 for the most extreme case, confirming that the gas is predominantly carried in bubbles whose rise triggers no significant convection as the detachment frequency is low.
 
    Corresponding results for the gas-evolution efficiency, $f_G$, are presented in figure \ref{fig:fG_Theta}($a$). As expected, $f_G$ increases significantly with fractional bubble coverage, $\Theta$. It approaches unity at lower currents and for the tightest spacings, consistent with the observations in figures \ref{fig:Gas_profile_Th}($a$) and \ref{fig:Sh_Th}($c$). Furthermore, $f_G$, generally decreases at higher current densities because the more frequent detachment events drive an increasingly stronger convection. As a result, the bulk of the gas is transported in dissolved form at $\vert i \vert= 10^4~\mathrm{A/m^2}$ even at the highest coverage of $\Theta = 0.56$. When comparing our data to the empirical relation provided by \cite{Vogt2011b}, it is important to keep in mind that in practice increasing current density generally leads to higher $\Theta$. To identify realistic combinations of $i$ and $\Theta$ in the simulations, we compare the parameter space to the $\Theta(i)$-relation given by \citet{Vogt2005} in figure \ref{fig:fG_Theta}($c$). Simulations lying close to this line are marked with filled symbols in figures \ref{fig:fG_Theta}($a$-$c$). When considering these data points only, our results for $f_G$ in figure \ref{fig:fG_Theta}($a$) approximately agree with the empirical relation for $\vert i \vert \sim 10^3~A/m^2$, but differences arise for higher and in particular for low current densities $\vert i \vert \leq 10^2~\mathrm{A/m^2}$.

    The results for $f_G$ are replotted in figure \ref{fig:fG_Theta}($b$), but this time as a function of $\Theta$ since this is the practically more relevant form. It also allows for a comparison to the relations provided by \citet{Vogt2011b,Vogt2013,Vogt2015,Vogt2017} based on theoretical considerations (see dashed gray and green lines in figure \ref{fig:Gas_profile_Th}($b$)). An obvious difference is that empirical relations are independent of $i$, whereas the data at any given $\Theta$ exhibit a significant variation depending on the current density. This difference is significantly less prominent when considering only the `realistic' cases, which are also reasonably well approximated by the expression of \citet{Vogt2017}, at least up to $\Theta \approx 0.3$. 

    Figure \ref{fig:fG_Theta}($c$) also includes results for the hydrogen supersaturation on the electrode in the steady state, $\zeta_{\mathrm{H_2,e}}$, which are shown as colour contours interpolated between the simulation data  points. Remarkably, the `realistic' cases close to the relation of \citet{Vogt2005} are seen to cover a very wide range of  $\zeta_{\mathrm{H_2,e}} \approx$ 10 up to very high values exceeding $10^3$. It should be noted, however, that  for the latter cases, the boundary layers are very thin (see figure \ref{fig:Gas_profile_Th}($b$)), such that the effective supersaturation on the scale of the bubble will be significantly lower.

     As a final point, we plot the bubble lifetime, $\tau_c$, in figure \ref{fig:fG_Theta}($d$). To compensate for the $1/i$-dependence, which leads to variations in $\tau_c$ over 4 orders of magnitude, the data is premultiplied by $i$. For $f_G = \textrm{const.}$, all curves in the presented form would be expected to collapse onto a single line with linear dependence on $\Theta$ based on \eqref{eq:fG}. While the linear trend is approximately preserved for all but the highest current density, the variations in $f_G$ lead to an increase in $i\tau_c$ with $i$ that is most pronounced for the highest current densities. 

	\subsection{Relating the electrode mass transfer to the effective buoyancy driving}\label{sec:Sh_electrode}
    \noindent The goal of this section is to provide scaling relations for the mass transport at the electrode based on the relevant physical transport mechanism. Our results so far have already highlighted the relevance of the convective flow driven by the departing bubbles. There is an analogy between the present configuration and single-phase buoyancy-driven convection in the sense that the detaching bubbles resemble the plumes of buoyant liquid in the latter case \citep{Climent1999}. Analyses based on boundary layer theory for convective heat transfer along vertical plates yield the power-law dependence on the Rayleigh number $Ra^{m}$, where the exponent $m$ asymptotically varies form $1/4$ for laminar flows to $1/3$ for turbulent flows at high $Ra$ \citep{Churchill1975}. The same power laws have empirically been shown to be valid for the convective heat transfer over horizontal plates and in particular for single-phase free-convective mass transfer over upward-facing horizontal electrodes by \citet{Wragg1968}. Beyond the laminar regime featuring an exponent of $0.25$, these authors  provided the relation 
    \begin{align}\label{eq:Sh_wragg}
		Sh=0.16 \left(Gr Sc \right)^{0.33},
	\end{align}
    for the mass transport in the turbulent regime, where the Grashof number $\Gr$ captures the buoyancy driving and Schmidt number is given by $Sc=\nu/D$. For two-phase buoyancy-driven convection, $\Gr$ can be defined to account for the effective buoyancy provided by the bubbles according to
	\begin{align}\label{Gr}
		\Gr=\frac{gd_b^3}{\nu^2} \frac{\rho_L - \rho_e}{\rho_e}=\frac{gd_b^3}{\nu^2}\frac{\rho_L - [(1-\epsilon)\rho_L+\epsilon\rho_G]}{(1-\epsilon)\rho_L+\epsilon\rho_G},
	\end{align}
	where $\rho_L$ is the density of the bulk electrolyte, $\rho_e$ is the mixture density at the electrode surface, $\rho_G$ is the gas density and $\epsilon$ is the gas volume fraction. Considering  $\rho_G \ll \rho_L$ yields the simplified expression
	\begin{align}\label{eq:Gr}
		\Gr=\frac{gd_b^3}{\nu^2}\frac{\epsilon}{1-\epsilon}.
	\end{align}
    Based on the fact that a single bubble is contained in a box with base area $A_e$ and height $u_b \tau_c$, where  $u_b$ denotes the bubble rise velocity, the gas volume fraction $\epsilon$ can be related to the volumetric flow rate of the gas, $\dot{V}_G=V_b/\tau_c$, by \citep{Zuber1963}
	\begin{align}\label{eq:Volume_Fraction}
			\epsilon=\frac{\dot{V}_G}{A_e u_b}.
     \end{align}
    For all cases investigated here we find that $\epsilon \ll 1$.  Assuming Stokes drag for the bubbles with no-slip interface yields the terminal velocity 
	\begin{align}\label{eq:ub}
		u_b=\frac{1}{18}\frac{gd_b^2}{\nu}\frac{\rho_L - \rho_G}{\rho_L},
	\end{align}
    which along with \eqref{eq:fG} leads to the final expression for $\Gr$ as
    \begin{align}\label{eq:Gr_final}
      \Gr=18 f_G d_b \frac{i}{n_eF}\frac{\mathcal R T_0}{P_0\nu}.
    \end{align}
    The ratio of buoyancy to viscous forces therefore depends linearly on the input parameters $d_b$, $f_G$, and in particular on $i$.  Consequently, the experimentally reported scaling of $Sh \sim i^{1/3}$ \citep{Janssen1973,Janssen1978,Janssen1979,Whitney1988} is equivalent to $Sh \sim \Gr^{1/3}$, provided that the product of the other two parameters ($d_b$ and $f_G$) in \eqref{eq:Gr_final} ($d_bf_G$) remains approximately constant with $i$.

    \begin{figure}
		\centering
		\includegraphics[width=1\columnwidth]{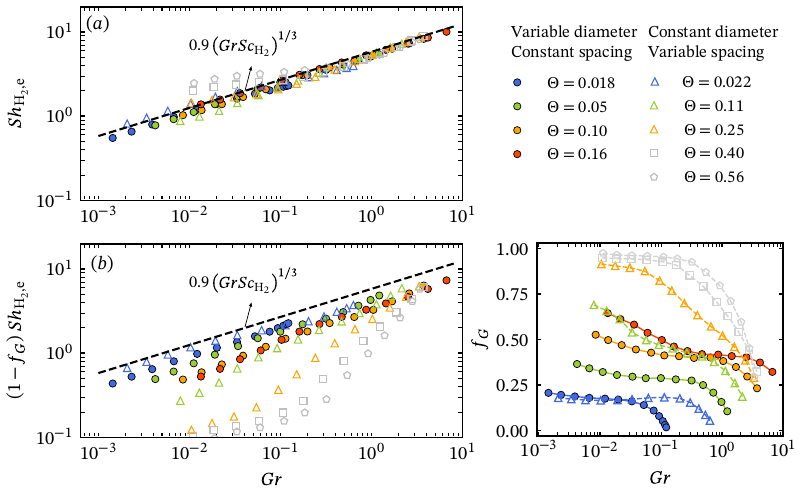}
		\caption{($a$) Sherwood number of hydrogen transport, $\Shge$ \eqref{eq:Sh_electrode}, averaged over one bubble lifetime in the statistically steady state, versus $\Gr$ for all cases studied in this work. ($b$) Fractional Sherwood number of hydrogen transport as dissolved gas in the liquid phase, $(1-f_G)\Shge$. ($c$) Corresponding values of  $f_G$  vs $\Gr$.}
		\label{fig:ShGr1}
	\end{figure}
 	 \begin{figure}
		\centering
		\includegraphics[width=1\columnwidth]{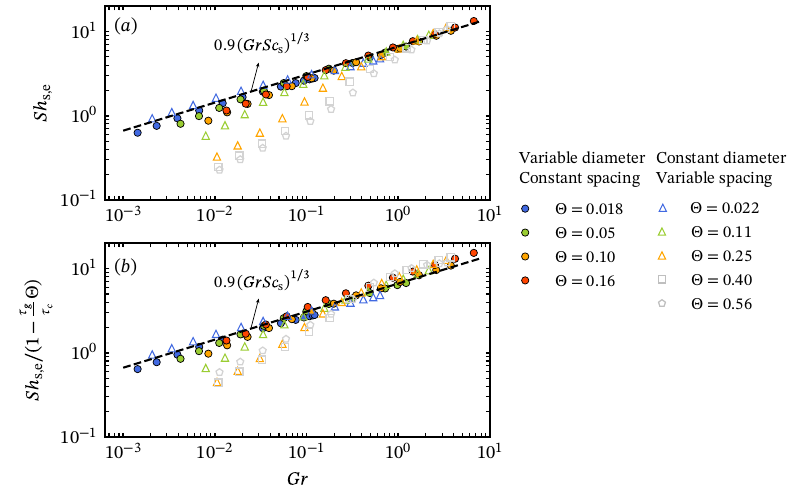}
		\caption{($a$) Sherwood number of electrolyte transport, $\Shs$ \eqref{eq:Sh_electrode}, averaged over one bubble lifetime in the statistically steady state, versus $\Gr$ \eqref{Gr} for all cases studied in this work. ($b$) $\Shs$ compensated for net blockage effect, $\Theta \tau_g/\tau_c$, caused by bubbles adhering to the electrode surface in the residence time. The legend specifies cases simulated for different bubble diameter and spacing using the corresponding fractional bubble coverage of the electrode, $\Theta$. The broken lines indicate the fitted power law, $\Shs = 1.0 \left(\Gr \Scs\right)^{1/3}$, in which $\Scs=\nu/D_s$. }
		\label{fig:ShGr2}
	\end{figure}
	Next, we consider the dependence of the Sherwood numbers for the mass transport at the electrode as a function of $\Gr$.  Figure \ref{fig:ShGr1}($a$) presents a plot of $\Shge$ vs. $\Gr$ for all data presented in $\S$ \ref{sec:Bub_size} and $\S$ \ref{sec:Bub_space}. In this form, the results very convincingly collapse onto a single line indicating the power law of $\Shge \sim Gr^{1/3}$, which supports the adoption of the single phase concept to the present configuration. Remarkably, the `turbulent' scaling exponent of 1/3 applies to the full range of $\Gr$ studied here, even though the flow is relatively weak and only intermittent in some cases (see figures \ref{fig:Gas_profile} and \ref{fig:Gas_profile_Th}). The data in figure \ref{fig:ShGr1}($a$) is well described by the fit
     \begin{align}\label{eq:Sh_SimFit}
		\Shge=0.9\left(\Gr \Scg \right)^{1/3},
	\end{align}
     where the difference in the prefactor compared to the single-phase equivalent \eqref{eq:Sh_wragg} is related to the multiphase nature of the present flow but also to the fact that a different length scale of bubble diameter is used here instead of lateral length scale of the electrode by \citet{Wragg1968}. The only significant deviation from \eqref{eq:Sh_SimFit} occurs for the `slow' (in terms of $\tau_c$) cases featuring a high $f_G$, for which the gas transport (carried almost exclusively inside the bubbles) is more efficient than buoyancy driving would suggest.

    It is important to note that here $\Shge$ and therefore \eqref{eq:Sh_SimFit} accounts for both the transport of gaseous and dissolved hydrogen. We can focus on the dissolved transport specifically by multiplying $\Shge$ with $(1-f_G)$, as is done in figure \ref{fig:ShGr1}($b$). For reference, a plot of $f_G$ for all data vs. $\Gr$ is also included in figure \ref{fig:ShGr1}($c$). Consistent with the fact that there is a wide spread in $f_G$ at any given $\Gr$, there is no collapse of the data in figure \ref{fig:ShGr1}($b$) underlining, that the analogy between single and multiphase buoyancy driven flows is applicable at the level of the total transport only. 

	The transport of the electrolyte, which entirely acts as a passive scalar here, for the most part falls in line with the trends discussed for $\Shge$. In particular, $\Shs$ primarily follows the power law of $\Shs \sim \Gr^{1/3}$ even with the same prefactor when accounting for the difference in $\mathit{Sc}$ as shown in figure \ref{fig:ShGr2}($a$). However, in accordance with figures \ref{fig:Sh_db}($c$) and \ref{fig:Sh_Th}($c$), $\Shs$ drops below this scaling at low $\Gr$ and high $\Theta$. This means that electrolyte transport from the bulk to the electrode surface is limited when the bubbles highly cover the electrode surface and adhere to it for a long period during their lifetime. According to \citet{Vogt1989,Vogt2012} a factor contributing to the lower transport of the electrolyte is the blockage effect due to the presence of the bubble as can be seen from the snapshots in figure \ref{fig:Cs_profile_Th}($a$).  To account for this, we divide $\Shs$ by the factor $\left( 1- \Theta \tau_g/\tau_c\right)$ in figure \ref{fig:ShGr2}($b$). Here, $1-\Theta$ is the fraction of the electrode not covered by the bubble and the additional timescale ratio accounts for the fact that the blockage applies only during the growth time $\tau_g$. Introducing this correction in fact reduces the deviations at lower $\Gr$ somewhat (but not fully) and the effect may therefore be relevant in this regime. However, the data for $\Gr \gtrapprox 1$ is overcompensated. In summary, it therefore appears that the fact that no sustained convection exists at high bubble coverages if $\Gr$ is low plays the most important role leading to the lower electrolyte transport. This leads to limitation in the applicability of the single-phase analogy for this case.  Nevertheless, it is worth noticing that the agreement with the 1/3 scaling law is much better for $\Shs$ (figure \ref{fig:ShGr2}($a$)) than for dissolved $\hyd$ (figure \ref{fig:ShGr1}($b$)), even though transport is exclusively within the electrolyte in both cases.
	
	\section{Mass transfer to the bubble}\label{sec:Sh_bub}

	\subsection{Bubble growth regimes}\label{sec:grwoth_scale}
	We now consider the dynamics of bubble growth and mass transfer into the bubble in more detail. The growth of the electrolytically generated gas bubbles can be approximated by an effective power law of
    \begin{align}\label{eq:growth_law}
        R(t)=\mathcal B t^x.
    \end{align}
    The limiting cases are as follows: During the very initial stage, when the growth of the bubble is strongly influenced by the inertia forces from the liquid \citep{Slooten1984}, an exponent of $x=1$ has been reported \citep{Westerheide1961,Glas1964,Brandon1985,Bashkatov2022}. For later times, depending on whether the bubble growth is limited by the diffusive mass transfer of dissolved gas to the interface \citep{Epstein1950,Scriven1959,Westerheide1961} or by the gas production rate in the reaction \citep{Darby1973,Verhaart1980,Higuera2022,Yang2015,Bashkatov2022}, exponents of $x=1/2$ or $x=1/3$ have been identified, respectively. However, in general, the effective value of the exponent in \eqref{eq:growth_law} deviates from these values due to the interplay between inertia, diffusion, and reaction rates.

\begin{figure}
		\centering
		\includegraphics[width=1\columnwidth]{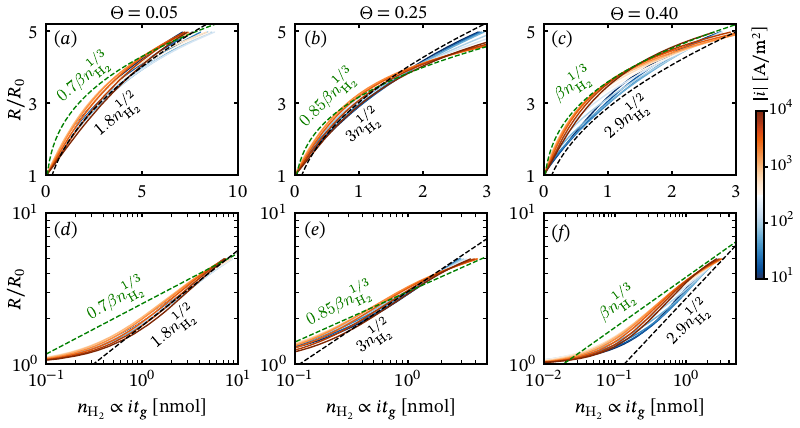}
		\caption{Temporal evolution of normalized bubble radius, $R/R_0$, versus the molar amount of hydrogen produced in the cathodic reaction, $n_\hyd=J_\hyd A_e t_g$, where $t_g$ is the time elapsed from the start of the bubble life in the stationary steady state. The results are for all the investigated current densities (distinguished with the colormap) at bubble coverages of $\Theta=0.05$ ($a$) $\Theta=0.25$, ($b$) and $\Theta=0.40$ ($c$). The second row ($d$-$e$) shows the same data as in ($a$-$c$) but with logarithmic scaling. The green and black broken lines show the power laws with exponents of $1/3$ and $1/2$, respectively. The prefactors for the 1/3 power law are adjusted  relative to the growth constant of purely reaction-limited bubble growth, $\beta=3.6~\mathrm{nmol^{-1/3}}$.}
		\label{fig:Growth_scale}
	\end{figure}

	Figure \ref{fig:Growth_scale} presents different growth dynamics in the statistically steady state, depending on current density and bubble coverage. Plotting the bubble radius versus the number of hydrogen moles, $n_\mathrm{H_2}$, produced in the reaction from the beginning of the bubble's lifetime, $t_g$, allows for easy comparison of the bubble growth dynamics over time for the full range of the current density. It is worth noting that $n_\mathrm{H_2}=J_{\mathrm{H_2}}A_e t_g$ and therefore $n_\mathrm{H_2} \sim i t_g$. Power laws with exponent $1/3$ and $1/2$ have been added for comparison in figure \ref{fig:Growth_scale} at different bubble coverages. Here, corrections are fitted to the prefactor $\beta=\left(3\mathcal R T_0/4\pi R_0^3 P_0 \right)^{1/3}$, which represents the value for purely reaction-limited growth (i.e. $f_G = 1$). For the lowest bubble coverage in figure \ref{fig:Growth_scale}($a$), the growth dynamics are best described by the exponent of $x=1/2$ at all current densities. This indicates that the rate of mass transfer to the bubble is controlled by the diffusive transfer of dissolved hydrogen to the bubble interface for these cases. However, a switch from $x=1/2$ to $1/3$ is appreciable as the current density increases at higher bubble coverages of $\Theta =0.25$ and 0.40 as presented in figure \ref{fig:Growth_scale}($b$) and \ref{fig:Growth_scale}($c$). At first sight, it may seem counter-intuitive that the reaction rate becomes more relevant as a limiting factor when it is increased. However, as discussed in the previous section, an increase in current density also significantly intensifies the convective transport which is then predominantly in the dissolved phase even at high $\Theta$. This reduces the boundary layer thickness and the amount of dissolved $\hyd$ (see figures \ref{fig:Gas_profile} and \ref{fig:Gas_profile_Th}), such that diffusive transport becomes increasingly less relevant compared to the faster reaction rate. Therefore, the exponent approaches $x=1/3$ and the prefactor approaches $\beta$, as observable form figure \ref{fig:Growth_scale}($b$) and \ref{fig:Growth_scale}($c$) where the bubble size evolution is better described by such power law at higher bubble coverages and current densities. 

	\subsection{Quantification of mass transport to the bubble }
\begin{figure}
		\centering
		\includegraphics[width=1\columnwidth]{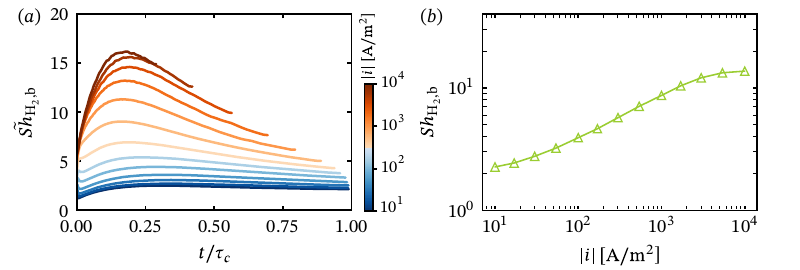}
		\caption{($a$) Temporal evolution of the Sherwood number for the bubble, $\Shgbt$ \eqref{eq:Sh_bubble}, during the entire bubble lifetime, $\tau_c$, in the statistically steady state and across the entire range of current density distinguished using the colormap. The data correspond to the case with bubble departure diameter of $d_b=0.5$ mm and a bubble spacing of $S=2$ mm, which leads to bubble coverage of $\Theta=0.1104$. ($b$) The corresponding averaged (over the bubble residence time $\tau_g$) Sherwood number of the bubble, $\Shgb$, over the residence time, $\tau_g$, plotted against the current density. }
		\label{fig:Shb_time}
	\end{figure}

    \noindent Figure \ref{fig:Shb_time}($a$) shows the transient behavior of $\Shgb$ according to \eqref{eq:Sh_bubble} over one bubble lifetime in the statistically steady state for varying current densities. Since bubble growth is neglected during the rise stage (see $\S$ \ref{sec:eq_dispersed} for further details) $\Shgb$ becomes equal to zero after the bubble break-off from the electrode surface. In figure \ref{fig:Shb_time}($a$), it can be observed that at low current densities, an equilibrated mass transfer rate to the bubble is established towards the end of the bubble residence time. This is evident from nearly constant values of $\Shgb$ at late stages of the growth phase, for current densities $\vert i \vert < 10^3~\mathrm{A/m^2}$ . In contrast, at higher current densities, $\Shgb$ remains in a transient all the way until the departure of the bubble. To study the mass transport to the bubble, the instantaneous $\Shgbt$, is averaged over the bubble residence time, $\tau_g$.  The corresponding results for the data presented in figure \ref{fig:Shb_time}($a$) are shown in figure \ref{fig:Shb_time}($b$) and indicate an increase of $\Shgb$  with increasing current density. 

\begin{figure}
		\centering
		\includegraphics[width=1\columnwidth]{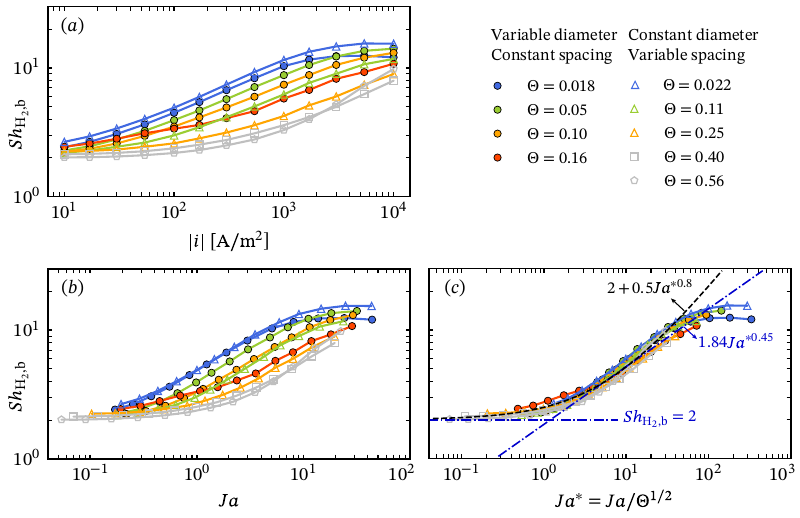}
		\caption{($a$) Sherwood number of hydrogen transport to the bubble,  $Sh_\mathrm{H_2,b}$, averaged over the bubble residence time, $\tau_g$, in the statistically steady state, as a function of the current density for all the simulation cases with varying bubble size or spacing. ($b$) $Sh_\mathrm{H_2,b}$ versus Jakob number, $Ja$, computed according to \eqref{eq:Ja}. ($c$) $Sh_\mathrm{H_2,b}$ versus $Ja^{\ast}$, i.e., the Jakob number corrected with $\Theta^{0.5} \approx d_b/S$ to account for the interference of mass transfer boundary layer on bubbles with each other. An approximate fit to the data and the two asymptotes are shown with black and blue broken lines respectively. The legend specifies cases simulated for different bubble diameter, $d_b$, and spacing, $S$, using the corresponding fractional bubble coverage of the electrode, $\Theta$. }
		\label{fig:ShbJa}
	\end{figure}

	To gain a broader understanding of hydrogen transport to the bubble and facilitate its quantification, we have plotted $\Shgb$ against current density in figure \ref{fig:ShbJa}($a$) for all the simulation cases, including those with variable bubble size or spacing. It is evident that at low current densities $\Shgb$ is nearly constant and then it starts to ramp up with current density at all of the simulated cases. Furthermore, the lower values of $\Shgb$ at higher $\Theta$ suggests that the normalized mass transfer to the bubble tends to decrease with bubble coverage.
	
	The current density is not directly related to the mass transfer into the bubble. In fact, the driving force for bubble growth is the concentration difference across the boundary layer developing at the bubble interface. The latter can be normalised with the gas concentration inside the bubble to yield the Jakob number, $\Ja$ \citep{Verhaart1980,Vogt1984_v2,Vogt2011a}
	\begin{align}\label{eq:Ja}
		\Ja=\frac{M_G}{\rho_G}\Delta C=\frac{\mathcal R T_0 }{P_0} \left(\Cg - \Csath \right),
	\end{align}
	where $M_G$ is the hydrogen molar mass and $\Cg$ is employed to estimate the concentration difference $\Delta C$ across the bubble boundary layer. At low $Ja \ll 1$ radial convection is negligible, such that $\Shgb$ remains constant. At moderate ($Ja \approx 1$) values and beyond, theoretical considerations suggest that the bubble Sherwood number becomes dependent only on $\Ja$, and no other parameter \citep{Epstein1950,Scriven1959,Verhaart1980,Vogt2011a}. However, the plot of $Sh_\mathrm{H_2,b}$ vs $\Ja$ for our results in figure \ref{fig:ShbJa}($b$) fails to collapse all the data onto a single curve. The reason for this is that in the theoretical considerations the effect of spatial confinement is not taken into account and the bubble is assumed to be in an infinitely large medium. However, especially for large $\Theta$ the growing bubbles interact and thereby enhance the effect of radial convection. This interaction becomes more prominent the smaller the bubble spacing $S$ is relative to the bubble diameter $d_b$. It therefore seems useful to define a compensated Jakob number $\Jas=\Ja/\Theta^{1/2}$ which additionally depends on the ratio $\Theta^{1/2} \approx d_b/S$. Figure \ref{fig:ShbJa}($c$) reports the results of $\Shgb$ versus the compensated Jakob number $\Jas$. Now a reasonable collapse of the data is achieved. An approximated fitting to the data gives
	\begin{align}\label{eq:Shb_Ja}
		\Shgb= 2 + 0.5\Jas^{0.8}.
	\end{align}
	It is shown as black broken line in figure \ref{fig:ShbJa}($c$). It is worth noting that for very low values of bubble coverage, particularly at $\Theta=0.018$ and $\Theta=0.022$, once again a nearly constant $\Shgb$ can be observed towards the upper limit of $\Ja^{\ast}$ (as shown in figure \ref{fig:ShbJa}($c$)) where deviation from \eqref{eq:Shb_Ja} occurs. This is related to the very short residence time of the bubble at very high current densities for these cases. As seen in the transient behaviour of $\Shgb(t)$ in figure \ref{fig:Shb_time}($a$), as the current density increases, the bubble departs from the electrode at increasingly earlier times before an equilibrated mass transfer to the bubble can be established. This leads to nearly constant averaged $\Shgb$ for such cases in figure \ref{fig:ShbJa}($c$), where a deviation from \eqref{eq:Shb_Ja} occurs.

	The relation \eqref{eq:Shb_Ja} for the mass transfer to the bubble is consistent with the classical theories of \citet{Epstein1950} and \citet{Scriven1959} for bubble growth in an infinitely large and uniformly supersaturated solution. The problem was later modified by \citet{Verhaart1980} to account for bubble growth over electrodes with non-uniform supersaturation around the bubble. The theories show a constant bubble Sherwood number of $\Shgb=2$ for small values of Jakob number, $\Ja \to 0$. Such condition is maintained in our simulations for high bubble coverages and low current densities where the concentration variation within the boundary layer is relatively low. The functional form used to represent the increase of $\Shgb$ for larger $\Jas$ in \eqref{eq:Shb_Ja} follows that suggested by \citet{Vogt2011a}, as approximation of the exact solution of \citet{Verhaart1980}. 
	
	It is useful to reformulate the definition of the Jakob number in terms of the Peclet number of mass transfer at the electrode, $\Pes$ (defined as the ratio of reaction to diffusion rates), and $\Shge$, as
	\begin{align}\label{eq:Ja2}
		\Jas=\frac{\Pes}{\Theta^{1/2}\Shge}, \quad \text{with}~\Pes=\frac{i}{2F}\frac{\mathcal R T_0}{P_0}\frac{d_b}{D_\hyd}.
	\end{align}
    Substituting the empirical fit \eqref{eq:Sh_SimFit} for $\Shge$, together with \eqref{eq:Shb_Ja}, leads to
	\begin{align}\label{eq:Shb_final1}
		\Shgb=2+0.5\left[\frac{\Pes}{\Theta^{1/2}~0.9\left(\Gr \Scg\right)^{1/3}}\right]^{0.8}.
	\end{align}
	The Grashof number can be expressed as $\Gr=18 f_G \Pes/\Scg$ (see \eqref{eq:Gr_final}) such that the final mass transfer relation for the bubble is given by
	\begin{align}\label{eq:Shb_final2}
		\Shgb=2+0.28\left(\frac{\Pes^{2/3}}{\Theta^{1/2} f_G^{1/3}}\right)^{0.8}.
	\end{align}
    Since $\Pes$ and $\Theta$ only depend on input parameters, the only previously unknown variable in \eqref{eq:Shb_final2}, just as in \eqref{eq:Sh_SimFit}, is the gas-evolution efficiency $f_G$. In order to enable a prediction solely based on input parameters, in the next section we will establish a suitable relation for $f_G$. 
	
	\section{Gas-evolution efficiency}\label{sec:fG}
     \noindent In steady-state conditions, we can restate the definition of the gas-evolution efficiency $f_G$ in \eqref{eq:fG} in terms of the cycle averaged molar fluxes into the bubble and at the electrode according to

	\begin{align}
		f_G = \frac{\bigintss_{0}^{\tau_c}\bigintss_{\partial V} D_{\mathrm{H_2}} \boldsymbol \nabla C_\hyd \cdot \hat{\mathbf n}_b \, \dd A_b \dd t}{\bigintss_{0}^{\tau_c} \bigintss_{A_e} D_\hyd \boldsymbol \nabla C_\hyd \cdot \hat{\mathbf n}_e \, \dd A_e \dd t} \sim \frac{D_\hyd \frac{ \left( \Cg -\Csath \right) }{\delta_b} d_b^2 }{D_\hyd \frac{\left(\Cg - C_{\mathrm{H_2},0}\right)}{\delta_e}A_e},
	\end{align}
	
	\noindent where  $\delta_b$ and $\delta_e$ are the boundary layer thickness normal to the bubble interface and electrode surface, respectively. Using $\Shgbs \sim  d_b/ \delta_b$, $\Shge \sim  d_b/ \delta_e$, $\Theta \sim d_b^2/A_e$  and noting that $ \left( \Cg - \Csath\right) / \left( \Cg - C_{\mathrm{H_2,0}}\right) \approx 1$, this leads to the expression
	\begin{align}\label{eq:fG_dimless}
		f_G= \alpha \Theta\frac{\Shgbs}{\Shge},
	\end{align}
    where the prefactor $\alpha$ is to be determined from the data. The difference between $\Shgb$ and $\Shgbs$ is that the former is averaged over the bubble residence time, $\tau_g$, whereas the latter is averaged over the entire bubble lifetime, $\tau_c$, consistent with the definition of $f_G$ \eqref{eq:fG}. Since bubble growth is disregarded during rise stage, $\Shgb(t)=0$ during this period such that the different definitions of the Sherwood numbers are related by  $\Shgbs=\left(\tau_g/\tau_c\right) \Shgb$. 
	
\begin{figure}
		\centering
		\includegraphics[width=1\columnwidth]{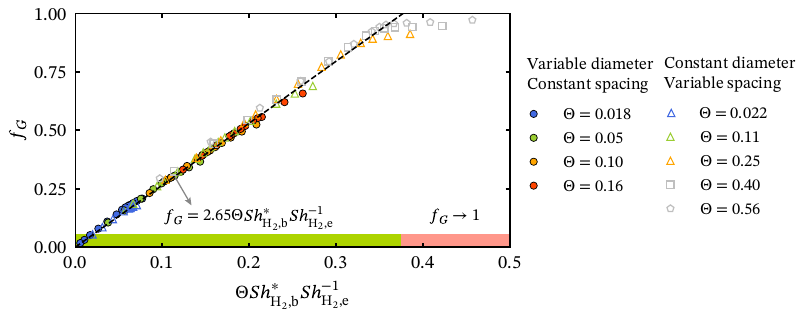}
		\caption{Gas-evolution efficiency, $f_G$, versus the dimensionless group $\Theta \Shgbs \Shgei$. The broken line shows the linear fit with slope $\alpha=2.65$ for $\Theta \Shgbs \Shgei < 0.375$, highlighted with green. For $\Theta \Shgbs \Shgei > 0.375$, highlighted with red, the gas-evolution efficiency approaches its upper bound, $f_G \to 1$.  }
		\label{fig:fG_group}
	\end{figure}

	Next, in figure \ref{fig:fG_group} the gas-evolution efficiency $f_G$ for all of the cases simulated here is plotted as a function of the dimensionless group, $\Theta \Shgbs \Shgei$. All data collapse onto a single line for $\Theta \Shgbs \Shgei < 0.375$, consistent with equation \eqref{eq:fG_dimless}. The slope is obtained as $\alpha=2.65$ based on the linear fit indicated as dashed line in the figure. For $\Theta \Shgbs \Shgei > 0.375$, the gas-evolution efficiency approaches its upper limit $f_G \to 1$ and the data level off close to this value.
    
    Inserting $\Shge$ from \eqref{eq:Sh_SimFit} and $\Shgb$ from \eqref{eq:Shb_final2} into \eqref{eq:fG_dimless} results in an implicit expression for $f_G$  that cannot be solved explicitly (see Appendix \ref{sec:app_expression}). Instead, we resort to piecewise solutions for $f_G$ by inserting the asymptotes of $\Shgb$ indicated by dashed blue lines in figure \ref{fig:ShbJa}($c$), into \eqref{eq:fG_dimless}. Doing so yields the explicit expressions 
	\begin{align}
		\label{eq:fG_exp1}
		&f_G=1.835\Theta^{3/4}\Pes^{-1/4}~~~~~, \quad &\text{for}~\Jas  \lessapprox 1,\\ 		
		&f_G=1.257\Theta^{0.522} \Pes^{-0.0225}, \quad &\text{for}~\Jas \gtrapprox 1.	
		\label{eq:fG_exp2}
	\end{align} 
	It should be noted that in the derivation of \eqref{eq:fG_exp1} and \eqref{eq:fG_exp2}, we have taken $\Shgb=\Shgbs$ presuming that $\tau_g/\tau_c\approx 1$, i.e. the bubble rise time is negligible. This is valid for our simulations at low and moderate current densities, whereas at high current densities $\tau_g$ ultimately becomes even smaller than the bubble rise time, $\tau_r$, violating this assumption (e.g. see figure \ref{fig:Gas_profile}($d$)). This can be considered an artifact of the simulations in which there is always a single bubble inside the computational box and the next bubble is initialized once the previous one has left the domain from the top boundary. Therefore the waiting time is equal to the bubble rise time, $\tau_r$, whereas experiments have revealed that the waiting time is extremely short especially at high current densities where the supersaturation level adjacent to the nucleation spot is very high \citep{Jones1999,Brussieux2011,Yang2015}. Therefore, the waiting time is insignificant and it can be safely considered that $\Shgbs=\Shgb$ for practical applications.

     For reference, in Appendix \ref{sec:app_expression} we have included explicit relations for $\Shge$ and $\Shgb$, resulting from combining \eqref{eq:fG_exp1} and \eqref{eq:fG_exp2} with \eqref{eq:Sh_SimFit} and \eqref{eq:Shb_final2}.
	
	\section{Further discussions and conclusions}\label{sec:conclusion}
	\noindent In this work, we set out to identify and quantify the governing mass transfer mechanism at gas-evolving electrodes by means of direct numerical simulations. Our work provides details on the mass transfer processes on a horizontal electrode subjected to successive growth and rise of electrolytically-generated gas bubbles. We employed the immersed boundary method to enforce the mass and momentum interfacial conditions on the bubble surface, and therefore, to solve for its growth rate as well as translational motion, employing Fick's law and particle equations of motion, respectively. To elucidate the main effects, we varied the current density within the range of $10 \leq \vert i \vert \leq 10^4~\mathrm{A/m^2}$ for different prescribed bubble sizes and spacings, expressed as fractional bubble coverage $\Theta$ of the electrode surface.   
	
	We quantified the cumulative hydrogen transport from the electrode surface (as dissolved gas and within the gas bubble) in figure \ref{fig:ShGr1} and that of electrolyte transport to the electrode in figure \ref{fig:ShGr2}. By drawing an analogy to single-phase heat and mass transfer problems, the buoyancy-driven convection induced by consecutively departing bubbles from the electrode surface was identified as the governing mass transfer mechanism. This finding was corroborated by a unique power law of $\Shj=0.9\left( \Gr \Scj\right)^{1/3}$, which was found to describe the hydrogen transport, and to a large part also the electrolyte transport at the electrode. For the electrolyte, a factor of $(1-\Theta)$ to compensate for the surface blockage effect reduces, yet does not fully eliminate, deviations from the power law at low $\Gr$. No such deviations occur at high $\Gr$, at which also most of the gas transport is in the dissolved state.

It is interesting to note that the observed 1/3-scaling implies that the dimensional mass flux is independent of the length-scale used in defining $Sh$ and $Gr$. Other definitions, such as utilizing the height of the domain as in \citet{Climent1999} or \citet{Lakkaraju2013}, are therefore just as valid. Nevertheless, the choice of $d_b$ is preferred here for consistency with earlier studies and practicality. Drawing on the corresponding heat-transfer regime in Rayleigh-Bénard (see e.g. \citet{Malkus1954,Grossmann2000}), the physical interpretation of this finding is that the mass transport is determined by the laminar boundary layers. The convective transport in the bulk, on the other hand, is so efficient that its details (such as the domain height) do not influence the overall rate. Reassuringly, this also implies independence of the domain height in our simulations, as is indeed observed (see Appendix \ref{sec:Height_effect}).
 
     Furthermore, we found a connection between the bubble growth dynamics and the hydrogen transport rate from the electrode. Specifically, as $\Gr$ increases with increasing current density and bubble coverage of the electrode, the growth dynamics of the bubble switch from diffusion-controlled, $R=\mathcal \alpha t^{1/2}$, to reaction-controlled, $R=\mathcal \beta t^{1/3}$, regime (see figure \ref{fig:Growth_scale}). This transition can be attributed to the high transport rate of hydrogen from the electrode surface at large $\Gr$ which prevailed over the gas production rate, thereby limiting the available oversaturation that would favour diffusive growth. Next, we quantified the hydrogen transport to the bubble as a function of the Jakob number $\Ja$. Our data showed no collapse when plotted against the conventional definition of $\Ja$. The agreement was much better, when additionally incorporating the ratio $d_b/S \sim \Theta^{1/2}$ into the definition of a modified Jakob number, $\Jas$, to account for the effect of neighbouring bubbles. With this modified definition, the resulting expression for mass transfer into the bubble is given by \eqref{eq:Shb_final2}. 
 
     Finally, we established a semi-empirical relation between the dimensionless mass transfer rates at the electrode and bubble interface and the gas-evolution efficiency $f_G$. Ultimately, this allowed us to provide explicit (i.e. depending on input parameters only) expressions for $f_G$ given by $\eqref{eq:fG_exp1}$ and $\eqref{eq:fG_exp2}$ and consequently also for the other response parameters $\Shge$ and $\Shgb$ (see Appendix \ref{sec:app_expression}). These findings can help quantify mass transfer rates in practical applications, provided typical bubble sizes and spacing on the electrode can be quantified.
	
	Our findings reveal different governing physics of mass transfer at gas-evolving electrodes than what was envisioned by \citet{Stephan1979,Vogt2011b,Vogt2015}, who attributed the rate-controlling mechanism of mass transfer to micro-processes induced by bubble growth and break-off from the electrode. As briefly introduced in $\S$ \ref{sec:Intro}, these micro-processes originate from three different sources: pure diffusion of fresh electrolyte to the electrode surface in the small region previously occupied by the bubble, convective flow induced by the expanding boundary of the bubble, and wake-flow after its break-off from the electrode. These processes impact the mass transfer in a microarea surrounding the nucleation spot whose size declines in time due to the bubble growth. For pure-diffusion transport of the reactant to the electrode during bubble growth, \citet{Vogt2015} modified the mass transfer relations established by \citet{Rousar1975}. To account for microconvection of bubble growth and break-off, they considered an analogy of the flow pattern around a growing bubble to lateral plug flow \citep{Stephan1979}, which was later modified with a boundary layer flow \citep{Vogt2015}. This approach allowed them to employ the mass transfer relations developed for such flows over a flat plate to quantify the averaged transport of reagent to the microarea within the time interval of bubble growth and break-off. They concluded that micro-processes in the small region surrounding the bubble were the rate-determining mechanism of mass transfer and prevailed over single-phase and two-phase free convection at moderate and high values of current density \citep{Vogt2011b}.
 
    Our results are inconsistent with these considerations due to several reasons. \citet{Vogt2015} assumed that the space previously occupied by the bubble was fully replenished with fresh electrolyte immediately after bubble break-off, and hence they employed Cottrell's relationship to predict the pure-diffusion mass transfer at the microarea. While this assumption holds true to some extent for high current densities, it is violated at low currents where the electrode boundary layer is much larger than the bubble break-off diameter (the bubble is fully immersed in the boundary layer, see figure \ref{fig:Gas_profile}). In such cases, stirring the solution in a region that is already depleted of reactant fails to fully replace the bubble volume with fresh bulk electrolyte. Likewise, the employed analogy to plug/boundary-layer flow over a flat plate is questionable because the predominantly wall-parallel advection of a depleted boundary layer caused by bubble growth hardly affects the wall-normal mass transfer. Consequently, we fail to observe enhanced mixing during growth periods in or simulations. 
	
	In contrast, our findings provide evidence that the flow pattern established by two-phase buoyancy-driven convection (see figure \ref{fig:Gas_profile}) is key in setting the mass transfer rate at the electrode. It is clearly visible from the $\hyd$ and $\acid$ concentration snapshots in figures \ref{fig:Gas_profile} and figure \ref{fig:Cs_proflie} that the concentration fields are changed in accordance with the flow pattern induced by bubble motion; i.e., an up-drought in bubble column, descent of the solution mixture between the bubbles and a roughly wall-parallel flow adjacent to the electrodes. Such flow pattern is analogous to those induced by plume emissions in single-phase free convection. In fact, the similarity of the mass transfer relations established in this work \eqref{eq:Sh_SimFit} to those of single-phase free convection \citep{Churchill1975,Wragg1968} proves that two-phase buoyancy-driven convection of departing bubbles is the rate-controlling mechanism of mass transfer at gas-evolving electrodes. This is further consistent with experimental measurements by \citet{Janssen1973,Janssen1978} and \citet{Janssen1979} where the thickness of the boundary layer on hydrogen-evolving electrodes followed the same power law as \eqref{eq:Sh_SimFit} when the bubble coalescence did not happen frequently. In summary, it therefore does not appear necessary to account for micro-processes, such as bubble growth, specifically when considering mass transfer.
	
	There remain some limitations that apply to this work. To avoid additional complications, we did not take into account the potential contribution of single-phase free convection, which arises from density gradients in the solution caused by concentration variations in the electrode and bubble boundary layers \citep{Ngamchuea2015,Novev2018}. Single-phase free convection might be of some influence at low current densities, where the bubbles adhere to the electrode for long period of time and allow the density gradients in the electrode boundary layer to develop to a sufficient extent necessary for triggering the instabilities \citep{Sepahi2022}. However, \citet{Sepahi2022} found that these instabilities are suppressed for bubble spacing of less than $\approx$ 2 mm, which is the case for most of the simulation cases here except those with the least bubble coverage of the electrode. At higher values of the current density where the frequency of bubble generation is relatively high, the induced flow of departing bubbles is very likely to suppress the single-phase free convection by reducing the density gradients in the cell or prevails over it if both mechanisms coexist. 
 
    The other neglected effect is the Marangoni convection \citep{Sternling1959,Yang2018,Park2023} arising from surface tension gradients along the interface due to the temperature increase or electrolyte depletion in bubbles proximity. Thermal Marangoni flow is mostly playing a role in electrolytically-generated gas bubbles on microelectrodes where the current density can easily surpass $10^6~\mathrm{A/m^2}$ in the bubble foot area and increase the temperature remarkably by ohmic heating \citep{Yang2018,Massing2019,Hossain2022,Bashkatov2023}. However, thermal Marangoni is likely less of a factor in the present configuration, as our current density does not exceed $10^4~\mathrm{A/m^2}$, which is not sufficient to increase the temperature considerably. However, solutal Marangoni as a result of electrolyte depletion \citep{Park2023} might play a role which needs further investigation in future works. Eventually, as our numerical solver treats the full 3-dimensional problem, we are able to extend this work to a set-up in which several bubbles are generated in a asymmetrical network of nucleation spots to study the collective effects of bubbles and replicate a system which mimics the relevant physics more accurately for practical applications.\\
	
	\noindent{\bf Acknowledgments\bf{.}} We thank Prof. Andrea Prosperetti for the fruitful discussions. This work was supported by the Netherlands Center for Multiscale Catalytic Energy Conversion (MCEC), an NWO Gravitation programme funded by the Ministry of Education, Culture and Science of the government of the Netherlands. This project has received funding from the European Research Council (ERC) under the European Union's Horizon 2020 research and innovation programme (grant agreement No. 950111, BU-PACT). This project has received funding from the European Union’s Horizon 2020 research and innovation programme under the Marie Skłodowska‐Curie (grant agreement No. 801359). We acknowledge PRACE for awarding us access to MareNostrum at Barcelona Supercomputing Center (BSC), Spain, and Irene at Trés Grand Centre de calcul du CEA (TGCC), France, under project No. 2021250115. \\

    \noindent{\bf Declaration of interest\bf{.}} The authors report no conflict of interest.\\
	
	\noindent{\bf Authors' ORCID\bf{.}} \\
	F. Sepahi \href{https://orcid.org/0000-0002-1581-2162}{orcid.org/0000-0002-1581-2162}\\
	R. Verzicco  \href{https://orcid.org/0000-0002-2690-9998}{orcid.org/0000-0002-2690-9998}.\\
	D. Lohse \href{https://orcid.org/0000-0003-4138-2255}{orcid.org/0000-0003-4138-2255}\\
	D. Krug \href{https://orcid.org/0000-0002-0627-5676}{orcid.org/0000-0002-0627-5676}.\\
	
	\appendix
	
	\section{Code verification}\label{sec:appendix}

 	\begin{figure}
		\centering
		\includegraphics[width=1\columnwidth]{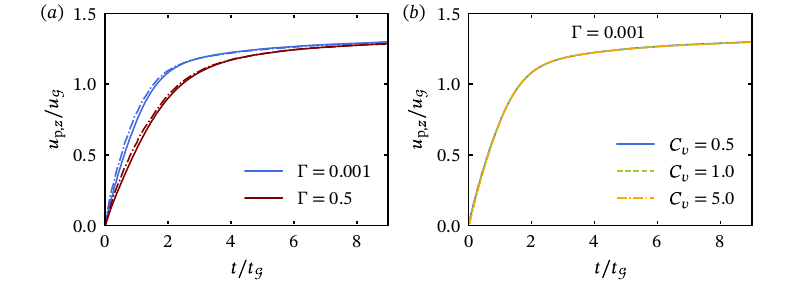}
		\caption{ ($a$) Temporal evolution of the normalized particle rise velocity for Galilei number $Ga=170$ at density ratios $\Gamma=0.001$ and 0.5, obtained from the present work (solid lines) and comparison to data from \citet{Schwarz2015} (broken lines). Virtual mass coefficients of $C_v=0.5$ and 0 have respectively been used for density ratios $\Gamma=0.001$ and 0.5. ($b$) Sensitivity of rise velocity to virtual mass coefficient for $\Gamma=0.001$.}
		\label{fig:VM}
	\end{figure}
 
	\subsection{Validation of bubble motion with IBM}\label{sec:virtual_mass}
	\noindent As discussed in $\S$ \ref{sec:eq_dispersed}, a remedy is required to solve for the bubble motion with IBM due to stability issues that arise at low gas to liquid density ratios. To mitigate this, the virtual mass approach  by \citet{Schwarz2015} is employed here and a virtual force, $\textbf{F}_v$ \eqref{eq:ub_param}, is added to both sides of \eqref{eq:ub}. To check the reliability of this method, we simulate the test case of \cite{Schwarz2015} using our code. The ascending motion of a light particle with a density ratios of $\Gamma=0.5$ and 0.001 in a quiescent viscous fluid is considered. Such flows are characterized by the  Galileo number defined as
	\begin{align}\label{eq:Galilei}
		Ga=\frac{\sqrt{\mid \Gamma - 1\mid g d_b^3}}{\nu}.
	\end{align}
	Additionally, the gravitational velocity and time scales read
	\begin{align}\label{eq:tg-ug}
		u_{\mathcal G}=\sqrt{\mid \Gamma - 1\mid g d_b} \quad \text{and} \quad t_{\mathcal G}=\sqrt{\frac{d_b}{\vert \Gamma - 1\vert g}},
	\end{align}
	respectively, and are utilized as reference values. The related parameters considered here are $Ga=170$, $g=\lVert{\mathbf{g}} \rVert=10$, $d_b=1$ and $\rho_L=1$. The size of the computational box is set to $\mathbf L = \left(6.4,6.4,12.8\right)d_b$ and is discretized with $\mathbf N = \left(256,256,512\right)$ cells in $x,y$ and $z$ directions, respectively. The sphere is initially at rest and released at $\mathbf{x}_{b,0}=\left( 3.2,3.2,0.6\right)d_b$. Periodic boundary conditions are applied in all directions and time marching is performed with steps of $\Delta t= 1\times 10^{-3}$ to exactly replicate the test case in \citet{Schwarz2015}. The simulation for $\Gamma=0.5$ is stable without modification of the original equation and is therefore run with $C_v=0$. Stability for $\Gamma=0.001$ is ensured by setting $C_v=0.5$. Figure \ref{fig:VM}($a$) presents the results for the time-evolution of the particle rise velocity $u_p$, along with the corresponding data from \cite{Schwarz2015}, with which excellent agreement is observed. Furthermore, we have performed the simulations for $\Gamma=0.001$ using different values of $C_v$ to check the sensitivity of results to the artificial virtual force. Figure \ref{fig:VM}($b$) shows that the particle rise velocity is quite insensitive to virtual mass. Hence, we conclude that this method can safely be employed to simulate the rising motion of electrolytically-generated gas bubbles with $\Gamma=0.001$ in this work.
	
	\subsection{Grid-independence check}\label{sec:grid_check}
\begin{figure}
		\centering
		\includegraphics[width=1\columnwidth]{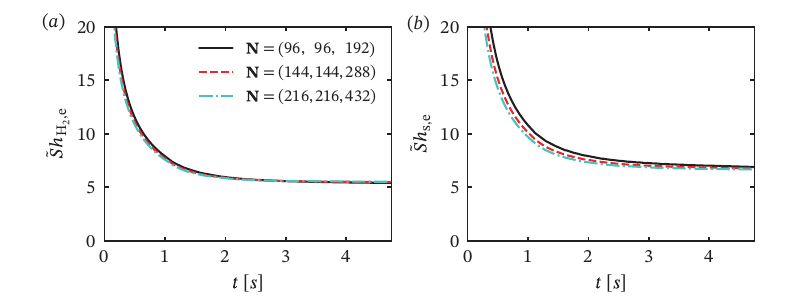}
		\caption{Grid independence check based on the on temporal evolution of $\hyd$ ($a$) and $\acid$ ($b$) Sherwood numbers on the electrode surface for the case presented in $\S$ \ref{sec:electrode}, i.e., $d_b=0.5$ mm and $S=2$ mm at the highest current density of $\vert i \vert=10^4~\mathrm{A/m^2}$. Base-grid sizes, introduced in panel (a), are refined by factor of 2 for $\hyd$ transport. Grid-independent results have been achieved for both species.}
		\label{fig:grid_check}
	\end{figure}

	\noindent To ensure the accuracy of the simulations, a grid-independence check has been performed on the case presented in $\S$ \ref{sec:electrode} with $d_b=0.5$ mm and $S=2$ mm. The highest current density of $\vert i \vert=10^4~\mathrm{A/m^2}$ featuring the thinnest boundary layer on the electrode (cf. figure \ref{fig:Gas_profile}) is selected for this purpose. Figure \ref{fig:grid_check}($a$) and \ref{fig:grid_check}($b$) show the time-evolution of the $\hyd$ and $\acid$ Sherwood numbers on the electrode surface for three grids with increasing resolution, confirming that the results are independent of the grid size in the investigated range. The base-grids are refined by a factor of two for $\hyd$ using a multiple resolution strategy as explained in $\S$ \ref{sec:eq_carrier}. This strategy ensures the hydrogen conservation in the system by sufficiently resolving the boundary layer thickness on the bubble interface (see Appendix \ref{sec:hyd_consv} ). Grid refinement is only applied for $\hyd$ transport, as dissolved hydrogen and its diffusion into the bubble determine the bubble dynamics and hence the whole hydrodynamics and mass transfer in the system. Based on the results in figure \ref{fig:grid_check}, the base grid resolution of $\mathbf N = \left(144,144,288 \right)$ is selected for the reference case and grid sizes for other cases with varying lateral size of the computational box have been adjusted  to keep the spatial resolution constant. This results in 36 grid cells across the bubble diameter if $d_b=0.5$ mm, whereas this value is 21 if $d_b=0.3$ mm.
	
	\subsection{Hydrogen conservation}\label{sec:hyd_consv}
\begin{figure}
		\centering
		\includegraphics[width=1\columnwidth]{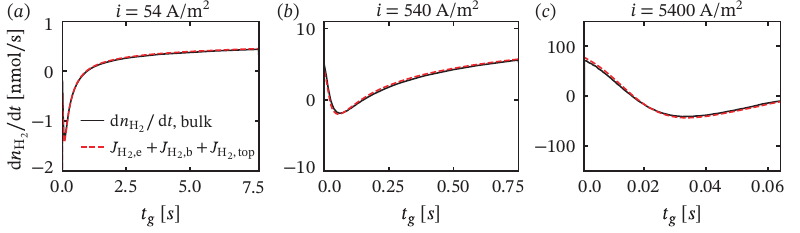}
		\caption{Hydrogen conservation check during the bubble residence time on the electrode at statistically steady state, performed for the case presented in $\S$ \ref{sec:electrode}, i.e.,  $d_b=0.5$ mm and $S=2$ mm at current densities $\vert i \vert=54$ ($a$) 540 ($b$) 5400 $\mathrm{A/m^2}$($c$). $t_g$ is the age of the bubble generated in the statistically steady state. Black solid lines are the rate of change of $\hyd$ moles in the solution mixture. Red broken lines are the summation of $\hyd$ production rate on the electrode ($J_{\mathrm{H_2,e}}$), desorption rate into the bubble ($J_{\mathrm{H_2,b}}$), and loss rate from the top boundary ($J_{\mathrm{H_2,top}}$). }
		\label{fig:hyd_consv}
	\end{figure}

     \noindent Obviously it is crucial to assure that the fluxes of dissolved hydrogen into the bubble interface, yielding the bubble growth rate, are accurately calculated with IBM, respecting mass conservation. To this end, we perform an analysis to check the conservation of hydrogen in the system.  This requires that the rate of change of $\hyd$ moles dissolved in the bulk electrolyte should be balanced with the net of $\hyd$ interfacial fluxes. The latter include the $\hyd$ production rate on the electrode surface  ($J_{\mathrm{H_2,e}}$), the desorption rate at the bubble interface ($J_{\mathrm{H_2,b}}$), and the outflux at the top boundary ($J_{\mathrm{H_2,top}}$). Figure \ref{fig:hyd_consv} compares the net interfacial fluxes with the rate of change of $\hyd$ in solution during bubble growth. This analysis concerns the reference case presented in $\S$ \ref{sec:electrode} ($d_b=0.5$ mm and $S=2$ mm) in the statistically steady state. It is evidenced by figure \ref{fig:hyd_consv} that our numerical scheme is conservative for hydrogen gas within the studied range of current density. However, higher current densities most likely demand finer spatial and temporal resolutions in order to capture the extremely thin mass boundary layers developed on the bubble and electrode interfaces. 
	
    \section{Additional expressions}\label{sec:app_expression}

    \noindent The implicit expression for the gas-evolution efficiency $f_G$, obtained by inserting \eqref{eq:Sh_SimFit} and \eqref{eq:Shb_final2} into \eqref{eq:fG_dimless}, reads

    \begin{align}
        f_G=1.12 \Theta \frac{2+ 0.28 \left( \frac{\Pes^{2/3}}{\Theta^{1/2} f_G ^{1/3}}\right)^{0.8}}{\left(f_G \Pes\right)^{1/3}},
    \end{align}

    \noindent which only has a piecewise solution. Inserting the expression for $f_G$ given by \eqref{eq:fG_exp1} and \eqref{eq:fG_exp2} into the fit of $\Shge$ given by \eqref{eq:Sh_SimFit} leads to an expression for $\Shge$ solely based on input parameters as

    \begin{align}
       &\Shge=2.89\left(\Theta \Pes \right)^{1/4}~~~~, \quad  &\text{for}~\Jas  \lessapprox 1,\\
       &\Shge=2.55 \Theta^{0.174} \Pes^{0.326}, \quad  &\text{for}~\Jas \gtrapprox 1.
    \end{align}

    \noindent Similarly, inserting \eqref{eq:fG_exp1} and \eqref{eq:fG_exp2} into \eqref{eq:Shb_final2} yields an expression for $\Shgb$ based on input parameters as 
    \begin{align}
       &\Shgb= 2+0.238\left(\frac{\Pes}{\Theta}\right)^{0.6}~, \quad  &\text{for}~\Jas  \lessapprox 1,\\
       &\Shgb=2+0.261 \left(\frac{\Pes}{\Theta} \right)^{0.54} , \quad  &\text{for}~\Jas \gtrapprox 1.
    \end{align}

    \section{Effect of domain height on electrode Sherwood numbers}\label{sec:Height_effect}
	
	\begin{figure}
		\centering
		\includegraphics[width=1\columnwidth]{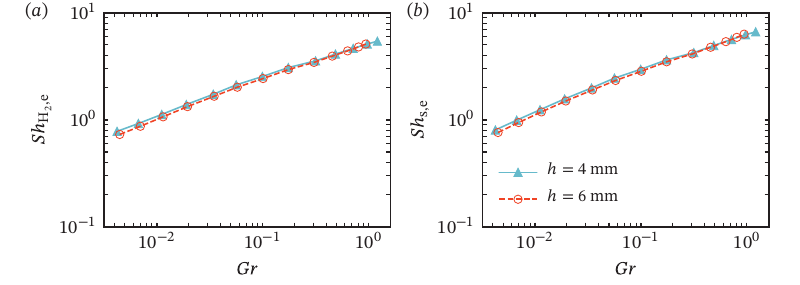}
		\caption{Sensitivity of the averaged Sherwood numbers of ($a$) hydrogen and ($b$) electrolyte transport at the electrode to the height of the computational domain.}
		\label{fig:Height_sensitivity}
	\end{figure}

	\bibliographystyle{jfm}
	\bibliography{jfm}
	
\end{document}